\documentclass[10pt,onecolumn,aps,prd,preprintnumbers,showpacs,superscriptaddress,nofootinbib,amsmath,amssymb,floats,floatfix,showkeys,notitlepage,longbibliography]{revtex4-1}

\usepackage{orcidlink}
\usepackage{comment}
\usepackage{lipsum}
\usepackage{graphicx}
\usepackage{subfigure}
\usepackage{palatino}
\usepackage{sans}
\usepackage{hyperref}
\hypersetup{colorlinks=true,linkcolor=blue,urlcolor=blue,citecolor=blue}
\usepackage[toc,page]{appendix}
\usepackage[normalem]{ulem}
\usepackage{adjustbox}
\usepackage{latexsym}
\usepackage{amsmath}
\usepackage{amssymb}
\usepackage{amsfonts}
\usepackage{dcolumn}
\usepackage{bm}
\usepackage{tikz}
\usetikzlibrary{decorations.pathmorphing}
\usepackage{bigints}
\usepackage{array,tabularx,multirow,booktabs}
\usepackage[tracking=true]{microtype}
\usepackage{soul} %for highlighting
\SetTracking{}{500}
\SetTracking{encoding={*}, shape=sc}{40}
\UseRawInputEncoding %for inputenc error%
\allowdisplaybreaks

\usepackage[utf8]{inputenc}
\usepackage{algorithm}
\usepackage{algorithmicx}
\usepackage{algpseudocode}
\usepackage{xcolor} % For colored text if needed

\begin{document} \sloppy
	
	\title{Interacting Tomonaga-Lüttinger liquid with impurity for interaction constant $K = 1/2$: Thermopower investigation, entropy variation and heat capacity densities associated with thermoelectric particle transport}
	
	\author{Abdellah Touati \orcidlink{0000-0003-4478-2529}} 
	\email{touati.abph@gmail.com, abdellah.touati@univ-bouira.dz}
	\affiliation{Department of Physics, Faculty of Exact Sciences, University of Bouira, Algeria.}
	\affiliation{Annex of the Higher Normal School of Bouzareah, University of Bouira, Algeria.}
	\author{Redoune Zamoum \orcidlink{0009-0002-5818-2989}
	} 
	\email{r.zamoum@univ-bouira.dz}
	\affiliation{Laboratory of Materials Physics and Optoelectronic Components,  Department of Physics, Faculty of Exact Sciences, University of Bouira, Algeria.}
	
	\author{Farhi Houssam eddine}
	\email{farhihoussameddine@gmail.com}
	\affiliation{Department of Physics, Faculty of Exact Sciences, University of Bouira, Algeria.}
	\begin{abstract}
		We investigate thermoelectric and thermodynamic properties of a Tomonaga-Luttinger liquid with interaction parameter $K=1/2$ in the presence of a localized impurity. Using bosonization and refermionization, we derive an exact expression for the thermopower in the nonlinear regime and obtain the corresponding Seebeck coefficient in linear response. At low temperature, the Seebeck coefficient follows a Mott-like relation governed by the energy dependence of the transmission coefficient and is found to be closely related to the entropy per particle. This connection allows us to derive using Kelvin formula the thermopower contribution to the entropy density variation and the charge-carriers contribution to the heat capacity density. Both quantities exhibit Mott-like behavior in the low-temperature regime. We apply our results to a fractional quantum Hall quantum point contact and to a one-channel quantum conductor coupled to an Ohmic environment. Our results demonstrate the close connection between thermoelectric transport and thermodynamic properties in interacting one-dimensional quantum systems.
	\end{abstract}
	
	%\pacs{04.70.Bw, 04.50.Kd, 04.25.-g, 95.30.Sf}
	
	\maketitle
	\tableofcontents
	\section{Introduction}\label{intro}
	In quantum systems, thermoelectricity is attracting increasing interest due to recent advances in quantum technologies. In this field, thermopower is one of the important physical quantities, as it is sensitive not only to charge transport but also to entropy, particle–hole asymmetry, many-body correlations, and the quantum states of the system. Thermopower provides information about which electrons carry the current and how their energy is distributed around the chemical potential. In the linear-response regime, thermopower is referred to as the Seebeck coefficient, which is sensitive to how the transmission varies around the Fermi energy. \
	
	Numerous experimental studies have aimed to measure the Seebeck coefficient in various systems, including quantum dots \cite{scheibner2005}, atomic-size contacts \cite{ludoph1999}, spin valves \cite{bakker2010}, nanowires \cite{hochbaum2008,boukai2008}, carbon nanotubes \cite{sumanasekera2002,small2003}, magnetic tunnel junctions \cite{walter2011}, and Kondo quantum dots \cite{scheibner2005}. It has been found that, in some materials exhibiting strong electronic interactions, the thermopower can be large \cite{terasaki1997,wang2003,foo2004}, thereby showing an enhancement under certain conditions \cite{haerter2006,peterson2007}. This behavior was observed in Ref. \cite{crepieux2010} for a metal-dot-metal junction under a time-dependent voltage using the Keldysh formalism. Furthermore, several theoretical studies have focused on the calculation of thermopower, including quantum dots in the coherent regime \cite{nakanishi2007}, superconducting quantum point contact (QPC) under a temperature difference using the electron scattering matrix \cite{pershoguba2019}, and an Aharonov–Bohm interferometer operating as a heat engine in the nonlinear regime \cite{haack2021}. \
	
	One of the main features that distinguishes thermopower, particularly in the linear-response regime, is that it provides information about other physical quantities as well as certain properties of the studied system. This important role becomes particularly evident in transport phenomena. Indeed, electrical conductance measures the magnitude of the transmission, whereas thermopower measures the energy dependence of the transmission. In addition, thermopower indicates whether transport is dominated by electrons or holes. Thus, electrical conductance and thermopower are complementary quantities, making their simultaneous measurement particularly valuable. This complementarity is clearly illustrated by Mott's formula for the Seebeck coefficient \cite{jonson1980}. Originally, Mott's formula was derived for independent electrons interacting with impurities and phonons within an adiabatic approximation, but it has also been widely used to analyze thermopower measurements \cite{houten1992,dzurak1997,llaguno2004,scheibner2005,small2003}. The connection between the Seebeck coefficient and transport quantities can be extended to mixed differential conductances \cite{crepieux2014} and to mixed heat-charge noise through the figure of merit \cite{crepieux2016}. In fractional quantum Hall (FQH) liquids, thermopower has been used to investigate the existence of non-Abelian quasiparticles \cite{moore1991}, as well as to probe their statistics and measure their quantum dimension \cite{yang2009}. \
	
	One of the important physical quantities characterizing the thermodynamics of electronic states is the entropy per particle, $s=S/N$ \cite{varlamov2016,shubnyi2018,sukhenko2018,kulynych2022,tsaran2017,galperin2018,grassano2018}. The Seebeck coefficient is closely related to the entropy per particle. An experimental verification of this close relationship was reported in metals with heavy electron- and hole-like quasiparticles \cite{behnia2004}. The relation $S_{Seebeck}=s/e$ can be interpreted as the entropy transported per charge carrier \cite{goupil2011}. Some authors refer to $s$ as the differential entropy per particle through the Maxwell relation
	$s=\frac{\partial S}{\partial N}\Big|_T=-\frac{\partial \mu}{\partial T}\Big|_N$,
	where $\mu$ is the chemical potential and $N$ is the particle number. Consequently, the Kelvin formula for the Seebeck coefficient, $S_{Seebeck}=\frac{1}{e}\frac{\partial S}{\partial N}\Big|_T$, has been shown to hold qualitatively for strongly correlated systems \cite{silk2009,zlatic2007} and in the incoherent metallic regime of ruthenates \cite{mravlje2016}. The close correspondence between the Seebeck coefficient and the entropy per particle has also been demonstrated in other systems, including zigzag graphene ribbons at low temperature, where $S_{Seebeck}$ is interpreted as the transport entropy per charge \cite{cortes2023}. Another example is quantum Hall states in Corbino geometry, where $S_{Seebeck}$ measures the entropy of quasiparticles, since $S_{Seebeck}=S/e^{\ast}N$ near the quantum Hall plateau at low temperature \cite{barlas2012}. \
	
	From a thermodynamic point of view, the entropy per particle describes how much information or disorder is associated with each electron. Since it is an intensive quantity, it allows comparisons between nanosystems of different sizes. The entropy per particle can be studied experimentally, indirectly, using Maxwell relations by measuring the temperature derivative of the chemical potential, $\partial \mu/\partial T$ \cite{kuntsevich2015}. On the other hand, the total entropy is a state function that characterizes the number of microscopic quantum states compatible with the macroscopic state of the system. For nanosystems, the total entropy is useful for probing characteristics such as quantum degeneracy, correlations, and many-body effects. Entropy can also be used to investigate the heat capacity, which is an important thermodynamic quantity because it measures the amount of energy required to increase the temperature. The specific heat is directly related to the density of states, low-energy excitations, and collective modes. Several theoretical studies based on solving the Schrödinger equation, determining the energy spectrum, and calculating the partition function have been carried out to obtain thermodynamic functions, including entropy and heat capacity, for various systems, such as multi-quantum-well and quantum-wire structures under high magnetic fields \cite{oh1994}, quantum wires and quantum dots under magnetic fields \cite{oh1995}, parabolic quantum wires in tilted magnetic fields \cite{ibragimov2001}, $GaAs$ quantum dot with Gaussian potential confinement in the presence of magnetic fields \cite{boyacioglu2012}, asymmetric quantum dots under the influence of tilted magnetic fields and temperature \cite{khordad2015}, triangular quantum wires \cite{khordad2016}, two-electron quantum dots under magnetic fields and parabolic interactions \cite{alshorman2018}, two-dimensional quantum dots \cite{yepes2019}, and quantum-dot superlattice systems in magnetic fields \cite{babanli2022}. Other calculations of entropy for specific systems have also been performed, such as the fractional entropy of multichannel Kondo systems obtained from conductance-charge relations \cite{han2022} and the calculation of quantum impurity entropy at the edge of a one-dimensional superconductor \cite{kattel2026}. Experimentally, examples of entropy measurements include spin-$1/2$ entropy measurements of strongly coupled quantum dots \cite{child2022} and direct entropy measurements in mesoscopic quantum systems \cite{hartman2018}. Heat-capacity measurements have also been performed in different systems, including multiple-well structures \cite{gornik1985,wang1988,bayot1996}, quantum Hall systems using thermalization time in the second Landau level at filling factors $\nu=5/2$ and $\nu=7/2$ \cite{schmidt2017}, and superconductors \cite{wen2020}. \
	
	In this work, the chosen system is an intercting Tomonaga--Luttinger liquid (TLL) with interaction parameter $K=1/2$ and an impurity localized at $x=0$. The Hamiltonian of this system is exactly solvable through bosonization and refermionization procedures. We focus on the calculation of the thermopower $Q$ using the wave function of the charge carrier flowing along the TLL. The obtained expression for the thermopower in the nonlinear regime (nonlinear Seebeck coefficient) is exact and valid for all voltage and temperature regimes. In the low-temperature limit, $Q$ obeys a Mott-like formula in terms of the logarithmic derivative of the transmission coefficient. We then derive the corresponding Seebeck coefficient $S_{Seebeck}$ in the linear-response regime and show that $S_{Seebeck}$ exhibits a Mott formula at low temperature, which reflects the energy dependence of the transmission coefficient (electrical conductance) around the Fermi energy. We then show that the Seebeck coefficient $S_{Seebeck}$ is very close to the entropy per particle $s$ at low temperature and low Fermi energy $\varepsilon_F$. This close relationship allows us to use the Kelvin formula for $S_{Seebeck}$ to derive the variation of the entropy density $\Delta \mathcal{S}$, which corresponds to the thermopower contribution to the variation in entropy density. This result can be interpreted as the total contribution of all charge carriers to the entropy variation. Finally, we derive the thermoelectric transport heat capacity density $\mathcal{C}$. This result can be interpreted as the contribution of the charges transported through the TLL to the heat capacity in the presence of the impurity. Both $\Delta \mathcal{S}$ and $\mathcal{C}$ exhibit a Mott-like formula as a function of the transmission coefficient in the low-temperature limit. We apply these results to two different systems mapped onto a TLL with an impurity: a quantum point contact between an FQH edge state at filling factor $\nu=1/3$ and a normal metal with $\nu=1$, and a one-channel quantum conductor coupled to an Ohmic environment. The results show that thermopower increases in the strong-backscattering regime, while the analysis of $\Delta \mathcal{S}$ and $\mathcal{C}$ reveals two different behaviors: a stable behavior at low temperature and an unstable behavior with increasing temperature. \
	
	The paper is organized as follows. In Section \ref{sec:SDM}, we describe the two studied systems mapped onto a TLL with an impurity. We provide a brief overview of bosonization and refermionization for a TLL with an impurity, together with expressions for the wave functions and transmission amplitudes. Then, in Section \ref{sec:thermo}, we study thermopower in the nonlinear regime. We analyze its voltage and temperature dependence and derive expressions in the limiting cases. In Section \ref{sec:LRR}, we investigate the linear-response regime. After deriving the expression for $S_{Seebeck}$ and analyzing the different temperature regimes, we show that, at low temperature and low Fermi energy, $S_{Seebeck}$ is very close to the entropy per particle $s$. We then derive the expressions for the density of entropy variation related to thermoelectric transport $\Delta \mathcal{S}$ and the charge-carriers contribution to the thermoelectric transport heat capacity density $\mathcal{C}$. We analyze their temperature and Fermi-energy dependences and provide the limiting cases. Finally, we conclude in Section \ref{conc}.

	%---------------------------------------------------------------------------------------
	\section{System description and Model}\label{sec:SDM}
	%---------------------------------------------------------------------------------------
	
	The main goal is to study the thermopower $Q$ in the nonlinear regime, and then obtain the Seebeck coefficient in the linear response regime. The Kelvin formula allows to investigate variation of entropy and specific heat of two different systems that can be mapped differently onto the same interacting TLL with interaction constant $K = 1/2$ and impurity located at $x=0$ which play the role of a diffusing center. The density Hamiltonian of TLL system is given by the bosonized Hamiltonian density \cite{chamon1996}:
	\begin{equation}\label{hambos}
		\mathcal{H}_{TLL}=\frac{v_F}{4\pi}\Big[ (\partial_x \phi_+)^2+(\partial_x \phi_-)^2 \Big]+A \delta(x) e^{-i\omega_0t/2+\phi_-(x)}+h.c~.
	\end{equation}
	where the amplitude of the tunneling part denoted by $A$ which described by the considered system and $\omega_0=e^*V/\hbar$ with $e^*$ denotes the electric charge of the quasi-particle. The bosonic field $\phi_-$ describe the tunneling interaction part of the TLL system, while $\phi_+$ describes a free chiral mode unaffected by the tunneling, where this decoupled system was obtained after an orthogonal of the Hamiltonian density, for more detail see Refs. \cite{chamon1996,sandler1999}.
	
	It is worthy to note that, the particular value of the interaction parameter $K=1/2$ allows us to use refermionization technique \cite{chamon1996}, in which the bosonic fields $\phi_\pm(t,x)$ can be refermionized by the new fields defined by $\gamma_\pm(t,x)=\frac{1}{2\pi}e^{i\phi_\pm(t,x)}$, where this new fermion are defined as such satisfies the following anti-commutation relation $\left\{\gamma_\pm(t,x),\gamma_\pm(t,y)\right\}=\delta(x-y)$. In term of the Klein factor $f$, we are able to write the new fermion field in this form $\psi_\pm(t,x)=\gamma_\pm(t,x)f$ \cite{chamon1996}, where the non-vanishing anti-commutations relations that satisfies by the operators $\psi(t,x),\,\psi^\dagger(t,x),\,f$ are given by \cite{chamon1996}
	\small
	\begin{equation}\label{eq:ff}
		\left\{\psi(t,x),\psi^\dagger(t,y)\right\}=\delta(x-y),\,\left\{\psi(x),f\right\}=0,\, \left\{f,f\right\}=2.
	\end{equation}
	The above Hamiltonian density of TLL system after refermionization becomes:
	\begin{eqnarray}\label{hamilton3}
		\mathcal{H}_{TLL}=\mathcal{H}_++\bigg[\psi(t,x)^\dagger\left(-i\frac{\partial}{\partial x}-\frac{\omega_0}{v_F}\right)\psi_-(t,x)+\sqrt{2\pi}\delta(x)\bigg(A \psi_-(t,x) f+A f \psi_-^\dagger(t,x)\bigg)\bigg],
	\end{eqnarray}
	where $\mathcal{H}_+$ is the Hamiltonian density of the free fields.
	This Hamiltonian density is completely solvable, and unable us to perform a non-perturbative analysis of electric transport properties. The new independent chiral fermions $\psi_-$ are given by \cite{chamon1996}:
	\begin{subequations}
		\begin{eqnarray}
			\psi^{in}_-(t,x)&=&\frac{1}{\sqrt{2\pi v_F}}\int_{-\infty}^{+\infty} a_\omega e^{i(\omega+\omega_0)\frac{x}{v_F}-i\omega\,t}d\omega,\label{psi}\\
			\psi^{out}_-(t,x)&=&\frac{1}{\sqrt{2\pi v_F}}\int_{-\infty}^{+\infty}  b_\omega e^{i(\omega+\omega_0)\frac{x}{v_F}-i\omega\,t}d\omega,\label{psi2}
		\end{eqnarray}
	\end{subequations}
	where the operator $b_\omega=t(\omega)a_\omega+r(\omega)a^\dag_{-\omega}$ is described by a combination of the annihilation and creation operators, $a^\dag_\omega$ and $a_\omega$, which obey the following commutation relation $\{a_\omega,a^\dag_{\omega'}\}=\delta_{\omega,\omega'}$. The frequency-dependent transmission $t(\omega)$ and reflection $r(\omega)$ amplitudes are given by:
	\begin{equation}
		t(\omega)=\frac{ \omega}{ \omega+i(\tilde{A}/2)},~~~~~~~r(\omega)=\frac{i(\tilde{A}/2)}{ \omega+i(\tilde{A}/2)}~,
	\end{equation}
	where $\tilde{A}\propto A$, and the transmission coefficient is written as follow:
	\begin{eqnarray}\label{eq:Tw}
		\mathcal{T}(\omega)=|t(\omega)|^2=\frac{4(\hbar\omega)^2}{4(\hbar\omega)^2+\tilde{A}^2},
	\end{eqnarray} 
	Obviously, in the perfect effective transmission $\mathcal{T}(\omega)$ is identical to $1$ when $\tilde{A}=0$ whatever the frequency is.
	
	%-----------------------------------------------
	\subsection{One-channel coherent conductor}\label{OCC}
	%-----------------------------------------------

	The first system is one-channel coherent conductor in series with a dissipative environment $Z(\omega)$ modeled by a resistance $R$, and its expression is given by:
	\begin{equation}
		Z(\omega)=\frac{R}{1+i\omega RC}
	\end{equation}
	where $\omega$, and $C$ represents the frequency and the effective capacitance of which includes implicitly of the conductor. In which follow we use this notation $\omega_{RC}=(RC)^{-1}$.
	\begin{figure}[h]
		\centering
		\includegraphics[width=0.4\textwidth]{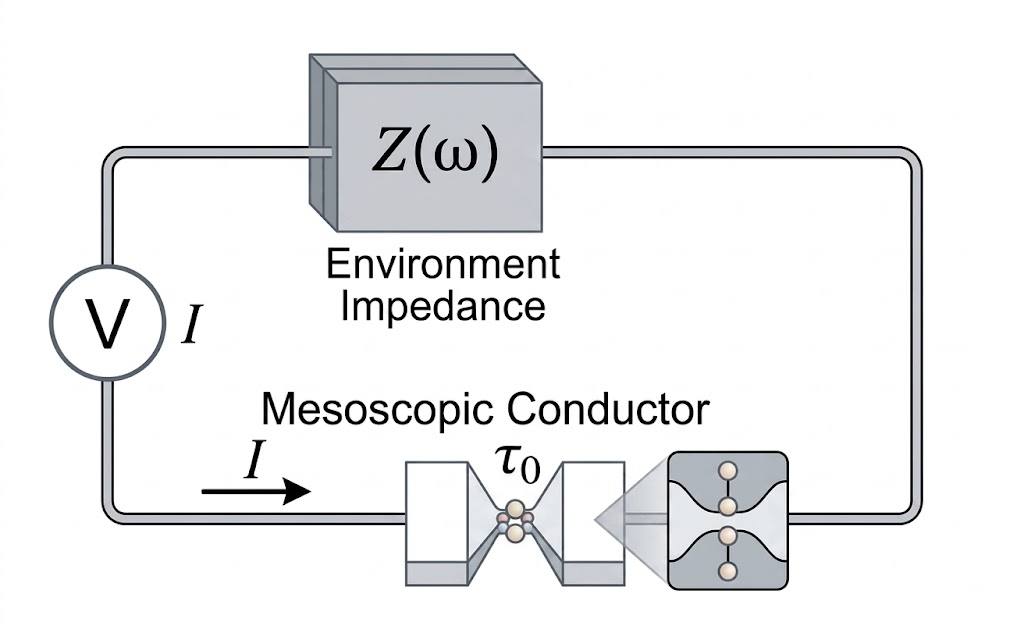}
		\caption{Schematic representation of a one-channel conductor with bare transmission $\tau_0$ embedded in an	electric circuit.}
		\label{fig:OCCS}
	\end{figure}
	
	The density Hamiltonian of this system is given by Eq. \eqref{hambos} substituting $A=\frac{\hbar\omega_F eV_B}{4\pi\sqrt{\pi}}$, where $eV_B=\hbar\omega_cv_B^{1/(1-K)}$ and $v_B$ is the effective backscattering amplitude, and is related with the effective transmission $\tau$ by this relation $v_B=\sqrt{(1-\tau)/\tau}$ and $\omega_c$ is the cut-off frequency \cite{zamoum2012}.

	Noting that the interacting constant $K$ in this system is related by the resistance and is given by the following equation \cite{safi2004}:
	\begin{equation}
		K=(1+R/R_q)^{-1}.
	\end{equation}
	For $R_q=h/e^2$ the quantum resistance, the interaction constante takes the value $K=1/2$. Now, the above Hamiltonian density coincides with that of the TLL given in Eq. \eqref{hambos}, and the TLL amplitude $\tilde{A}=eV_B/(2\hbar)$ in the transmission and reflection amplitudes leads to:
	\begin{equation}
		t(\omega)=\frac{\tau \omega}{ \tau\omega+i(1-\tau)\omega_c},~~~~~~~r(\omega)=\frac{i(1-\tau)\omega_c}{ \tau\omega+i(1-\tau)\omega_c}~,
	\end{equation}
	
	%-----------------------------------------------
	\subsection{Quantum point contact of two different FQH state}
	%-----------------------------------------------
	
	The second system is a hybrid normal metal $\nu=1$ linked with a $\nu=1/3$  FQH edge states at a point-like junction or QPC. At this point, a particle approaching from one edge can either be transmitted to the opposite edge or reflected back along its original path, thereby introducing both forward and backward scattering channels.
	\begin{figure}[h]
		\centering
		\includegraphics[width=0.35\textwidth]{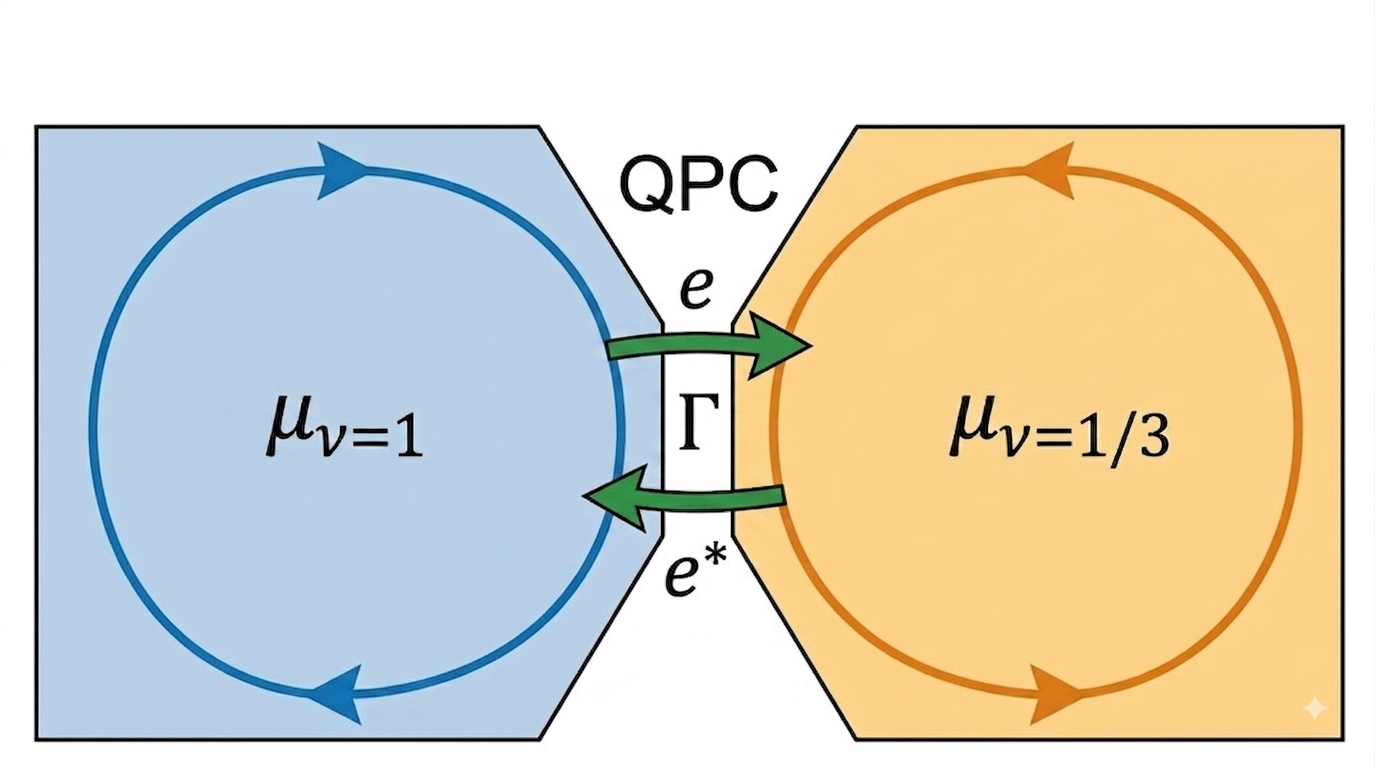}
		\caption{Schematic representation of a hybrid normal metal $\nu=1$ linked with a $\nu=1/3$ fractional quantum Hall edge states.}
		\label{fig:HFQHS}
	\end{figure}
	
	The density Hamiltonian of this system is given by the Eq. \eqref{hambos}, substituting $A=\Gamma$, where $\Gamma$ represent the the strength of the electron tunneling amplitude \cite{sandler1999}. The interacting constant $K$ for this system is related to the filling factor $\nu$ for each edges, and is given by:
	\begin{equation}
		K=\frac{2}{1+\nu^{-1}},
	\end{equation}
	where $1$ correspond to the hybrid metal and $\nu$ is for FQH system. This system can be mapped to an TLL system, when the filling factor of FQH system is $\nu=1/3$, it leads to the specific value of the interacting constant $K=1/2$. At this value the above Hamiltonian density become similar to the one given by Eq. \eqref{hambos}, so the refermionization techniques is applicable to this system, and the exact treatment of the electric transport is possible now. Then the transmission and reflection amplitudes in term of the new TLL amplitude $\hbar\tilde{A}=\Gamma_B\propto|\Gamma|^2$ are given by:
	\begin{equation}
		t(\omega)=\frac{ 2\hbar\omega}{ 2\hbar\omega+i\Gamma_B},~~~~~~~r(\omega)=\frac{i\Gamma_B}{ 2\hbar\omega+i\Gamma_B}~,
	\end{equation}
	
	These expressions fully describe the scattering behavior of the chiral fermions through the QPC. 
	%---------------------------------------------------------------------------------------
	%---------------------------------------------------------------------------------------

	%---------------------------------------------------------------------------------------
	\section{Thermoelectricity of TLL system with impurity}\label{sec:thermo}
	%---------------------------------------------------------------------------------------

	In what follows, we give a brief review of some quantities linked to electrical transport in TLL with impurity and $K=1/2$, such as the current, the conductance and the transport coefficients, in order to get the thermopower of these systems.
	
	As we see above both systems present the same solution and the principal difference is observed in the transmission and reflection amplitudes, and that unable us to find a more general current, conductance, and transport coefficients, by using the proper amplitude $\tilde{A}$ for each system that described above.

	%--------------------------------------------------------------------
	\subsection{Electric current and conductance}
	%--------------------------------------------------------------------

	The total current passing through a TLL with impurity and $K=1/2$ is given by \cite{zamoum2012,zamoum2013}:
	\begin{align}\label{ecurrent}
		I(V)&=\frac{e}{4\pi}\int_{-\infty}^{+\infty}\mathcal{T}(\omega)\bigg[f\left(\hbar\omega-\frac{eV}{2}\right)-f\left(\hbar\omega+\frac{eV}{2}\right)\bigg]d\omega, 
	\end{align}
	where $f(\hbar\omega)=\big(1+\exp(\hbar\omega/(k_BT))\big)^{-1}$ is the Fermi-Dirac distribution function. Noting that, the above expression of the electric current is exact and corresponds to the Landauer formulation of the current. Using the electrical current one can obtain the differential conductance, which is defined as $G=e\partial I/\partial \mu$ (where $\mu=eV/2$). Using the above expression we find:
	\begin{align}\label{econductance}
		G(V)&=\frac{e^2\beta}{4\pi}\sum_{\pm}\int_{-\infty}^{+\infty} \mathcal{T}(\omega)\bigg[f\left(\hbar\omega\pm\frac{eV}{2}\right)\bigg(1-f\left(\hbar\omega\pm\frac{eV}{2}\right)\bigg)\bigg]d\omega,
	\end{align}
	As we have mentioned in the beggining of the section, the expressions of the electrical current and conductance for each system can be obtain by using the proper amplitude $\tilde{A}$ in the transmission coefficient $\mathcal{T}(\omega)$.

	%--------------------------------------------------------------------
	\subsection{Thermopower} 
	%--------------------------------------------------------------------

	One of important materiel properties is the thermopower, which measures its ability to generate a voltage from a temperature difference. In the TLL theory the thermopower $Q$ is defined by the following relation \cite{kane1996}:
	\begin{equation}
		Q=\frac{\mathcal{L}_{12}}{\mathcal{L}_{11}}=\frac{1}{eT}\frac{\sum_{\pm}\int_{-\infty}^{+\infty} \mathcal{T}(\omega)\Bigg[\pm f\big(\hbar\omega\mp eV/2\big)\big(1-f\big(\hbar\omega\mp eV/2\big)\big)\big(\hbar\omega\mp eV/2\big)\Bigg]d\omega}{\sum_{\pm}\int_{-\infty}^{+\infty} \mathcal{T}(\omega)\bigg[f\left(\hbar\omega\pm\frac{eV}{2}\right)-f\left(\hbar\omega\pm\frac{eV}{2}\right)^2\bigg]d\omega}, \label{eq:Thermopower}
	\end{equation}
	where $\mathcal{L}_{11}=G(V)$ and $\mathcal{L}_{12}=\partial I(V)/\partial T$ are the diagonal and off-diagonal transport coefficients respectively. Now in order to analyze the behavior of the thermopower of each system we need to use the proper transmission coefficient $\mathcal{T}(\omega)$.

	\begin{figure}[!h]
		\begin{center}
			\includegraphics[width=0.32\textwidth]{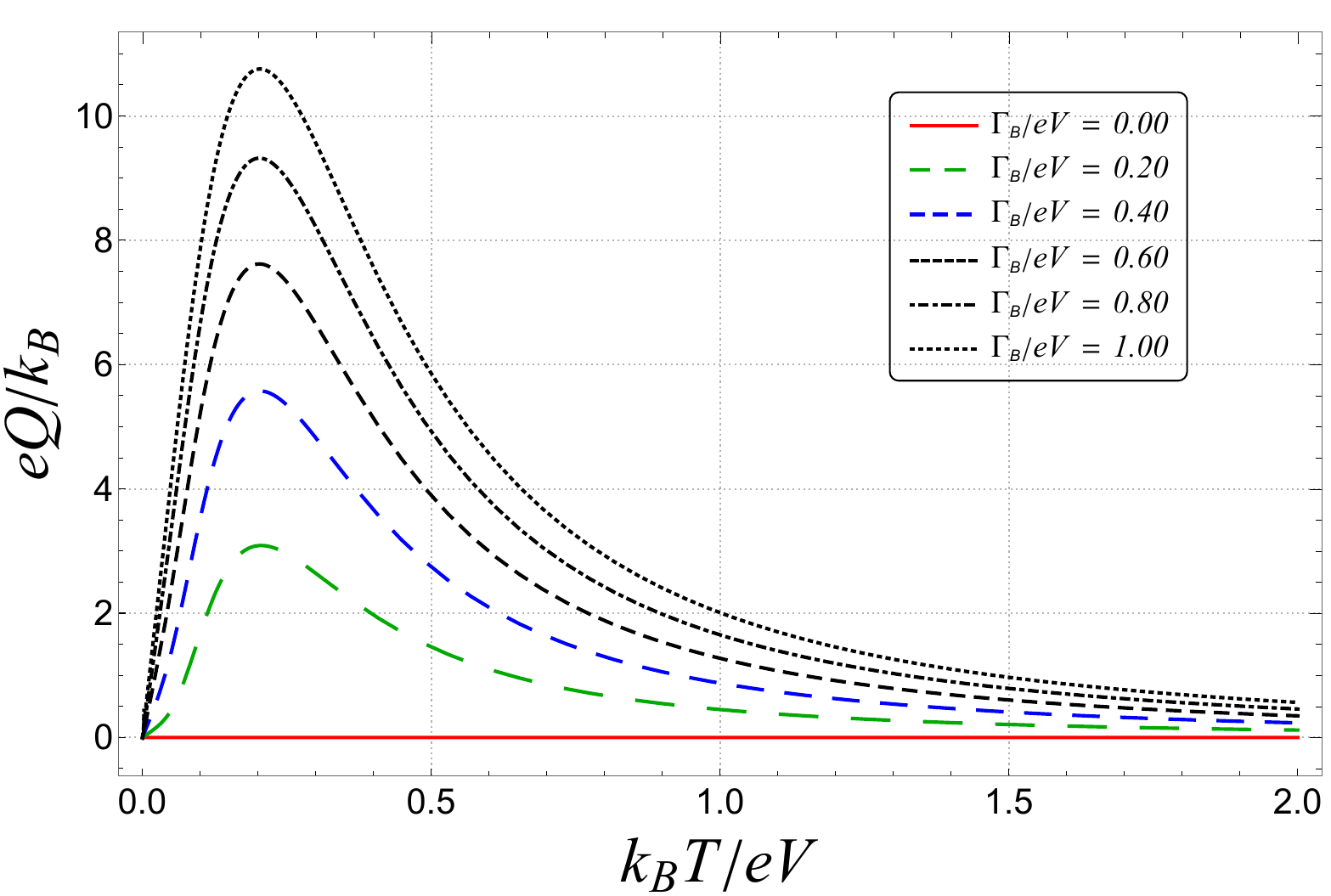}
			\includegraphics[width=0.32\textwidth]{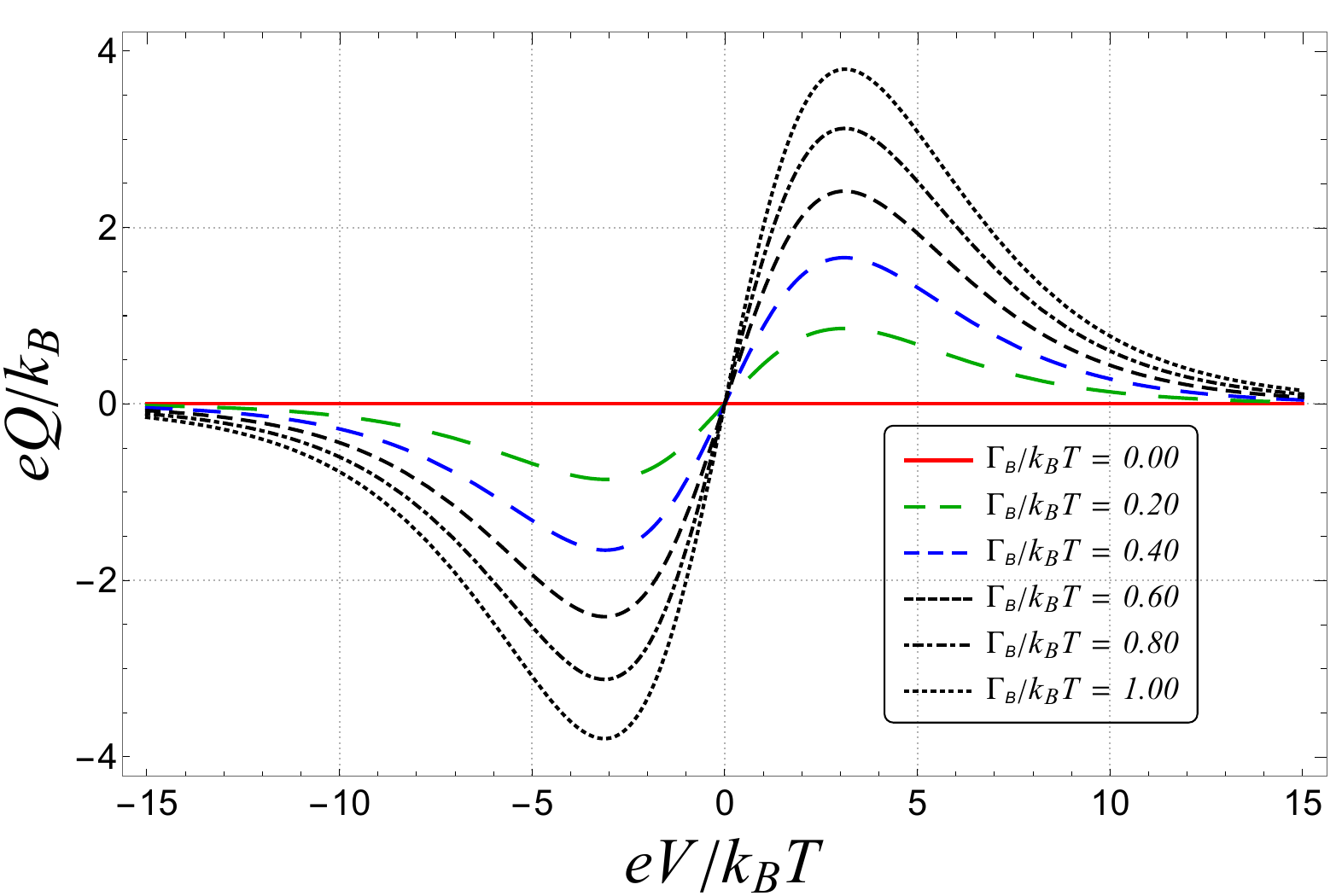}\\
			\includegraphics[width=0.32\textwidth]{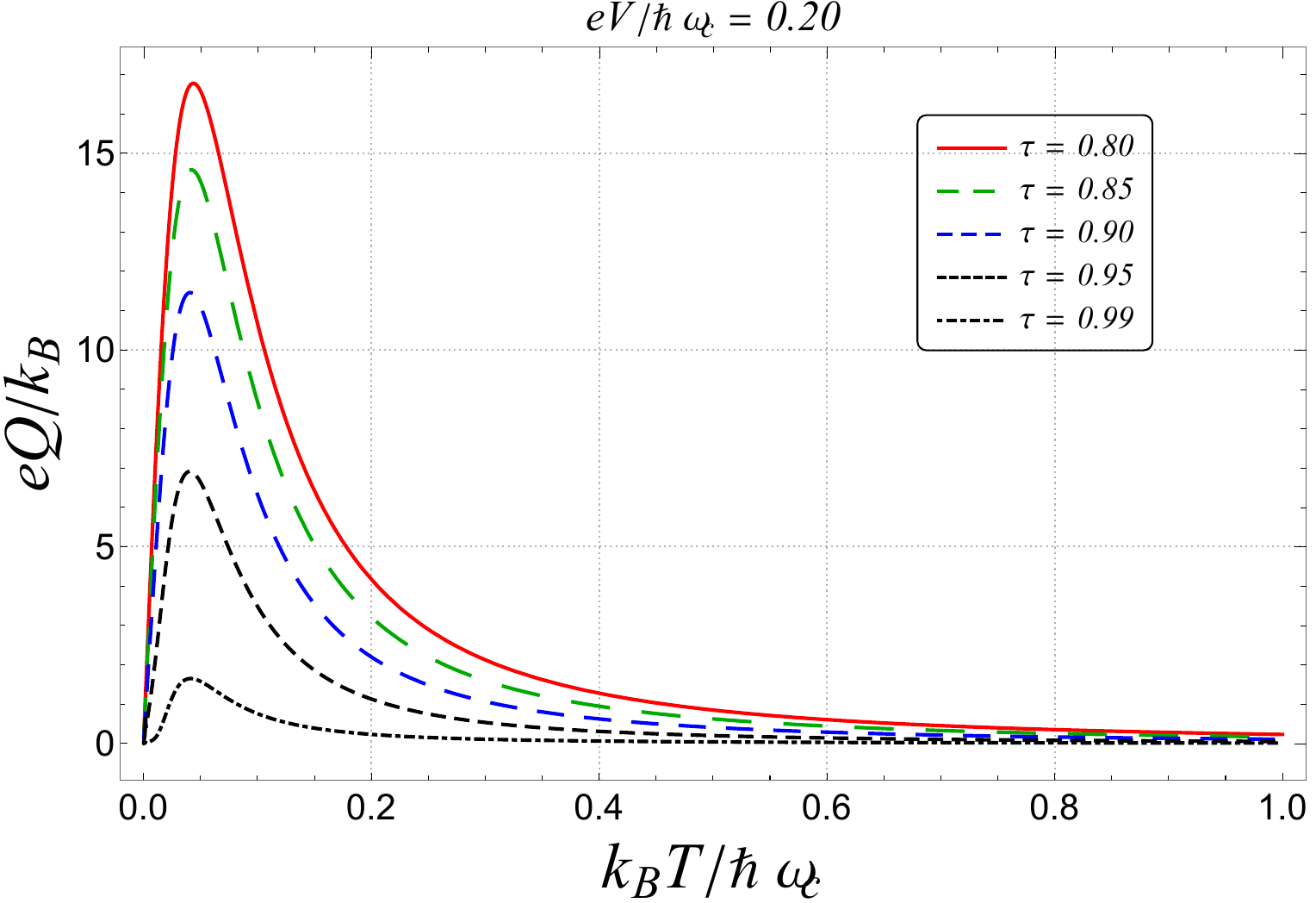}
			\includegraphics[width=0.32\textwidth]{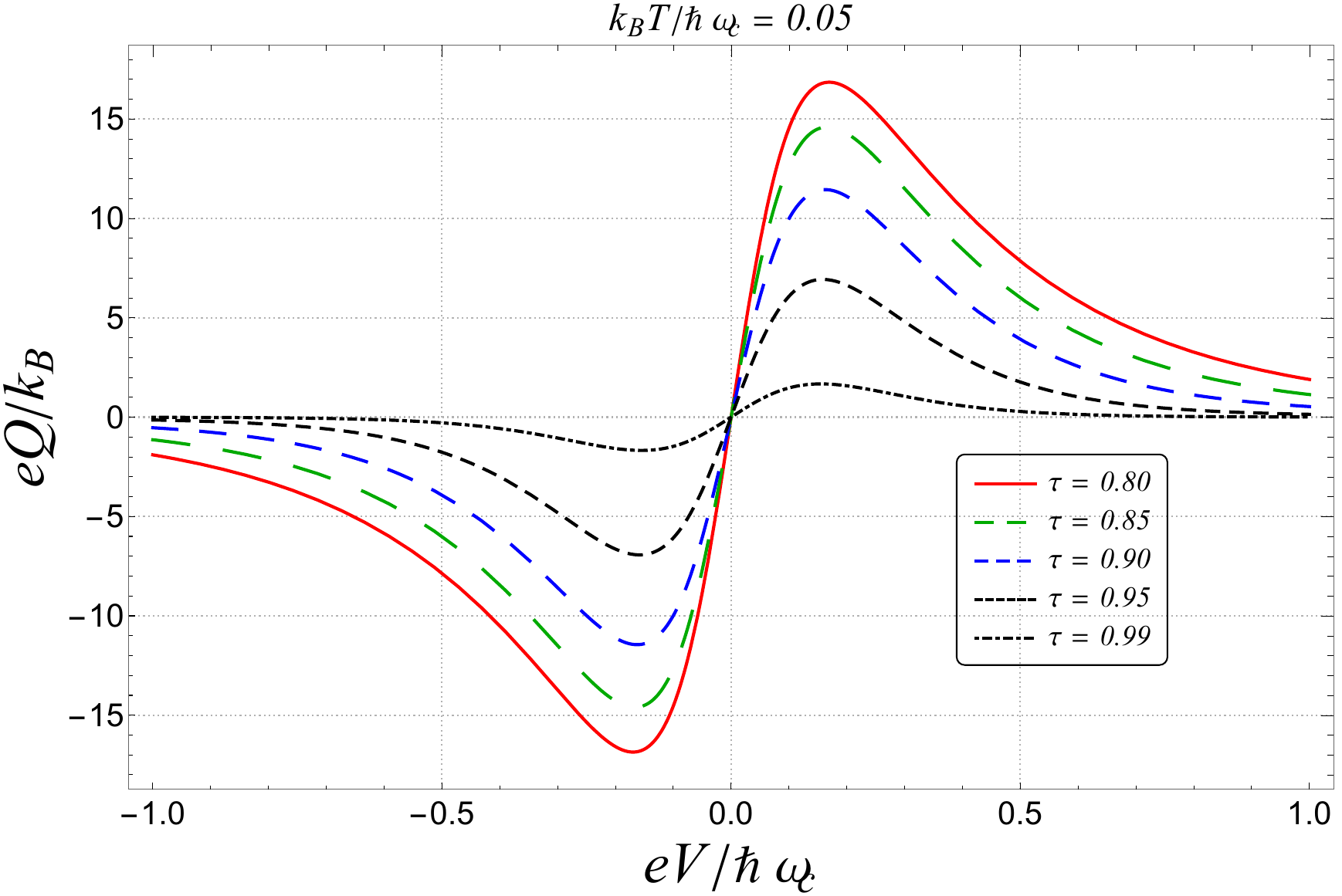}
			\caption{Variation of thermopower: In first row, versus temperature $k_BT/eV$ in left panel and applied voltage $eV/k_BT$ in right panel and for different backscattering amplitude $\Gamma_B/eV$, and $\Gamma_B/k_BT$ for QPC system.  In second row, versus temperature $k_BT/\hbar \omega_c$ in left panel and applied voltage $eV/\hbar \omega_c$ in right panel and for different effective transmission $\tau$, for OCC system.}\label{fig:QFQH-OCC}
		\end{center} 
	\end{figure}
	In Fig. \ref{fig:QFQH-OCC}, we plot the thermopower in units of $k_B/e$ versus temperature $k_bT$ (left pannels) and applied voltage $eV$ (right pannels). 
	\paragraph{Temperature dependence}
	For the QPC (high left pannel) or OCC (low left pannel), the Seebeck effect vanishes when $T \to 0$, no thermovoltage is genereted since there is an increasingly narrow energy window arround $\mu$. When $T$ increases, $Q$ increases and peaks. This is due to more energetic charge carriers contribution to thermopower, since $k_BT$ broads and captures the particule-hole asymmetry introduced by the impurity. The competition between incresing energy carrier number and broades energy spread yields a maximum in $Q$. At this point, $Q$ is sensitive to the transmission over an extanded energy intervalle $\mu-eV/2 \lesssim E \lesssim \mu+eV/2$ which corresponds to a finite transport window. The same behavior is confirmed in the reference \cite{trocha2025}, where the authors show that $S\approx0$ inside symmetry gap and large $|S|$ at higher $T$ once quasiparticle tunneling dominates for a quantum dot attached to normal metal and topological superconductor. At heigh $T$, the Fermi distribution of hot or cold carriers smear out over many levels and contributions above and below $\mu$ largely cancels. Thus, $Q$ deacreses after the peak. 
	\paragraph{Bias voltage dependence} 
	For the QPC (high right pannel) or OCC (low right pannel), at zero bias $Q(V=0)=0$ and the system is electron-hole symmetric. When $eV \ll k_BT$ the non linear corrections are weak. Increasing bias voltage defines an energy window $\Delta E \approx eV$ over which carrires can flow. This window acts as an energy filter. In fact, for increasing $V$ (or eventually $T$) transmission coefficient becomes mor energy dependent and varies accross the energy window, a large thermovoltage develops. The maximas arise when voltage (or temperature) becomes optimally matched to the energy scale over which the backscattering induced transmission changes: $eV,k_BT \sim E_{char}$. In the case of OCC $E_{char}=eV_B$ with voltage scale $V_B$ (see section \ref{OCC}). Here, the energy asymmetry is maximally resolved and $\vert Q \vert$ becomes large. For larger voltage, the bias window far exceeds variation in $\mathcal{T}(\omega)$, different energy contributions average each other and $Q$ decreases. 
	\paragraph{Role of backscattering strength $\Gamma_B$ and effective transmission $\tau$}
	In the case of QPC, higher $\Gamma_B$ makes the transmission more energy dependent. Thus, thermopower is enhanced, and the peak reaches a maximum value since carriers at specific energies dominate transport but reduces electrical conductance \cite{zamoum2012}, which is a thermoelectric trade-off. 
	For OCC, higher effective transmission $\tau$ tends to supress $Q$ beceause a very transparant channel has less energy asymmetry relative to $\mu$. In the auther hand, lower $\tau$ corresponding to strong backscattering, generates strong enhancement of energy filtring giving larger $Q$. A larger $Q$ requires a strong particule-hole asymmetry around $\mu$. This analysis is analogue to signature of a TLL with impurity connected to reservoirs where electron-electron interaction enhances thermopower \cite{krive2001}.  
	\paragraph{Signe reversal}
	The signe reversal observed in the right curves of figure \ref{fig:QFQH-OCC}, occures when the dominant carriers switch from electrons (energy above $\mu$) to holes (energy below $\mu$). At the particule-hole symmetry point $Q=0$. Varying $\mu$ or tunning $V$ shifts the energy window. The curves show an approximate linear inversion $Q(-V)=-Q(V)$ around the center point. The signe reversal is a robust indicator of energy asymmetry (can be calibrated against the polarity of experimental measurment).
	\paragraph{The cut-off effect}
	The energy cut-off $\hbar \omega_c$ in the case of OCC sets the upper limit of validity for effective description. For $k_BT,eV \ll \hbar \omega_c$, the system remains in the low energy TLL regime. When $k_BT$ or $eV$ become comparable to $\hbar \omega_c$ the peak is observed. For $k_BT \gtrsim \hbar \omega_c$ the curves deviate frome the low energy law: $Q$ decreases more rapidly signaling the saturation of the TLL effects, which is visible in the flattening the curves at values of $k_BT$ or $eV$ near $\hbar \omega_c$.
	
	\begin{figure}[!h]
		\begin{center}
			\includegraphics[width=0.32\textwidth]{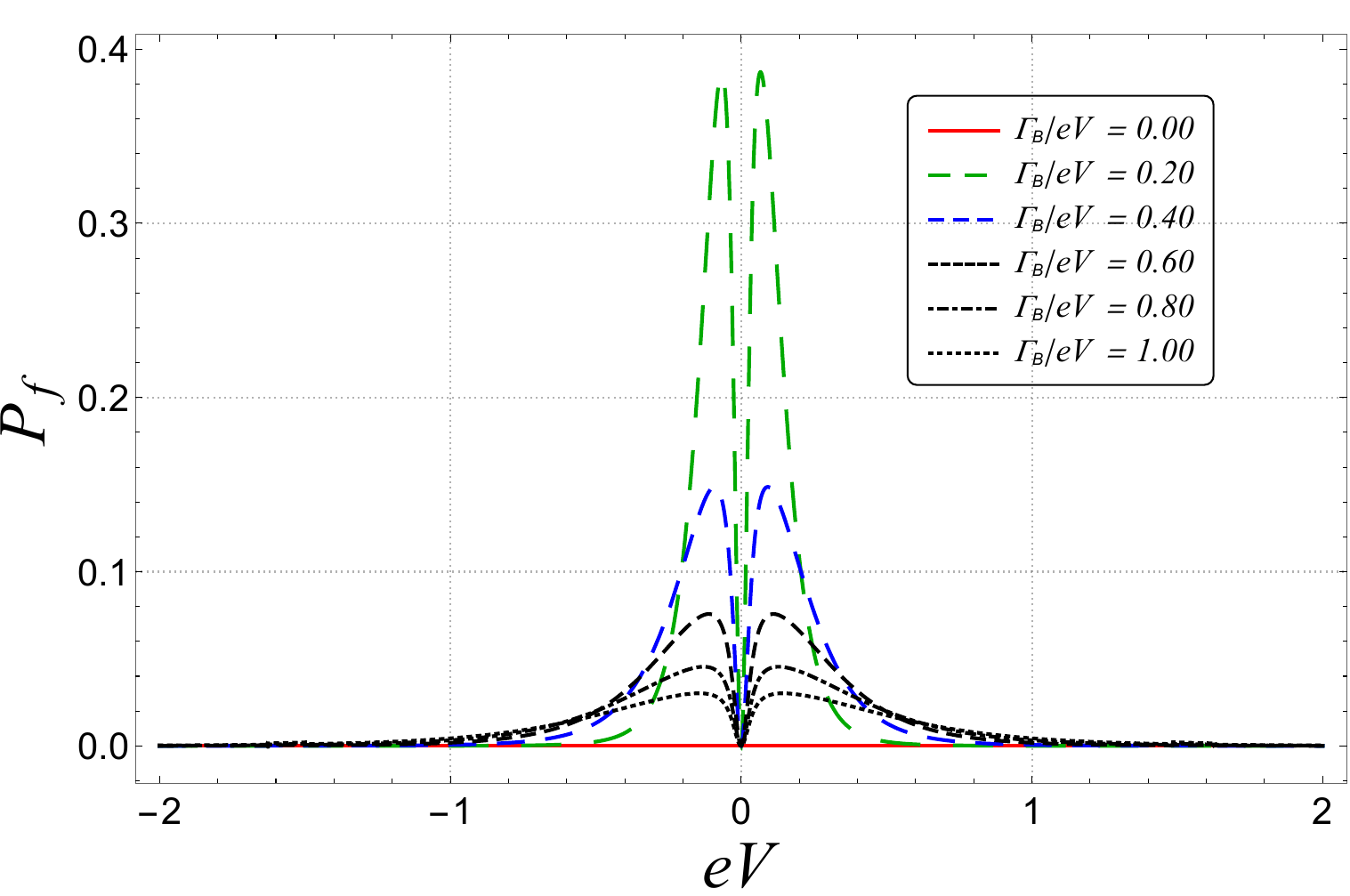}
			\includegraphics[width=0.32\textwidth]{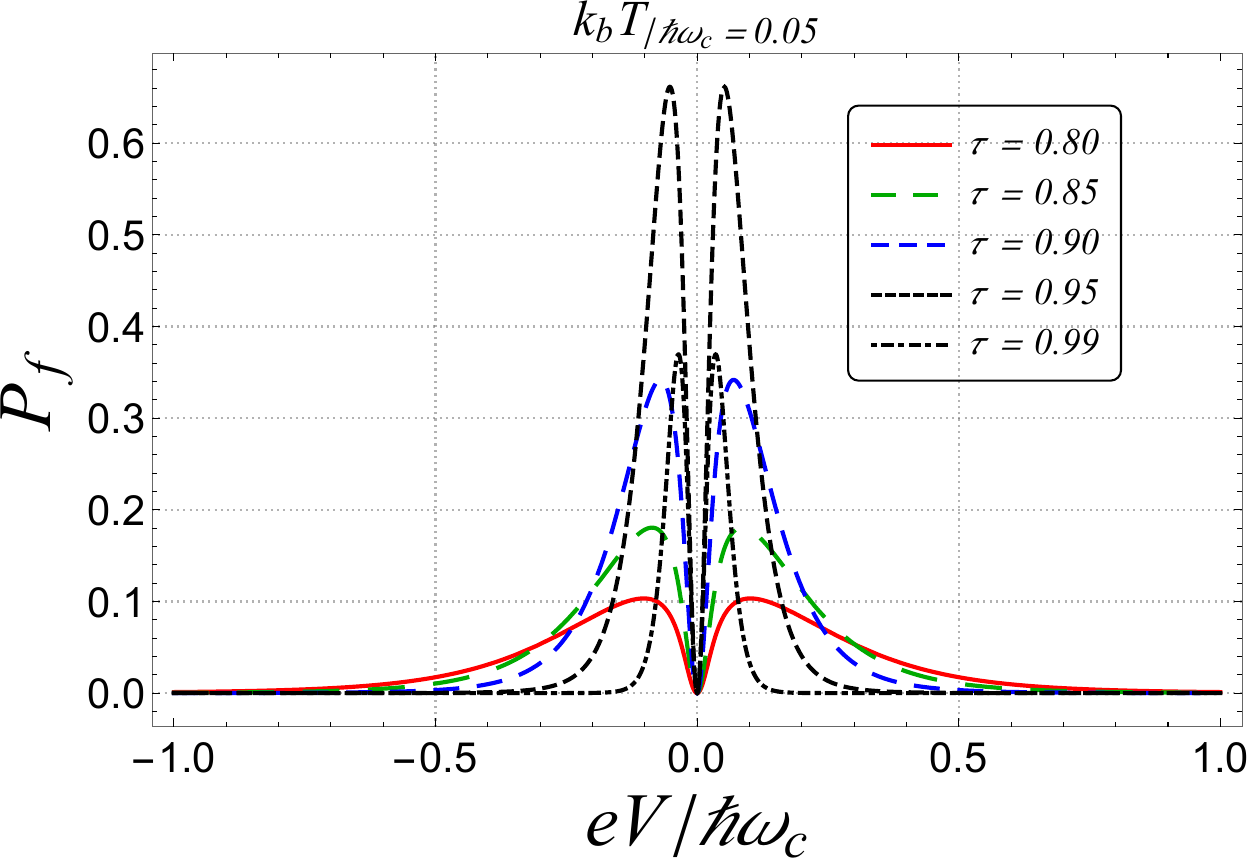}\\
			\caption{Power factor: In left pannel, versus bias voltage $eV/k_BT$ for different backscattering amplitude $\Gamma_B/k_BT$ for QPC system.  In right pannel,versus bias voltage $eV/\hbar \omega_c$ for different effective transmission $\tau$, for OCC system.}\label{fig:Pf}
		\end{center} 
	\end{figure}
	
	We plot in figure \ref{fig:Pf} the power factor $P_f=G Q^2$. As mantionned above, a trade-off between enhancing thermopower and supressing electrical conductance exsists. The curves clearly indicate an optimum backscatering regime where the power factor is maximal. This optimal area is located between weak backscatering and intermediate backscatering regimes for both systems the QPC and the OCC.
	
	\subsection{Temperature regimes behaviors}

	In what follows, we investigate the thermopower in TLL with impurity in different temperature regimes behaviors. In order to highlights central aspects such as the impact of thermal fluctuations, the mechanisms of energy transport, and the dependence on the transmission function, and backscattering amplitude, we go through the low and high temperature limit for the thermopower behavior, and we include the impact of weak and strong backscattering in the low temperature regime.

	\subsubsection{Low temperature limit:}
	
	In order to write the thermopower expression \eqref{eq:Thermopower} in the low temperature regime $k_BT\ll (\hbar\omega\mp eV/2)$, we use the Sommerfeld approximation for the derivative of the Fermi-Dirac distribution, which given by:
	\begin{equation}
		\int_{-\infty}^{+\infty}\! \mathrm{d}\varepsilon\; \varphi(\varepsilon)\,\bigg(-\frac{\partial f(\varepsilon)}{\partial \varepsilon }\bigg)
		\xrightarrow[]{T\to0}  \varphi(\mu)
		+ \frac{\pi^2}{6}\,\varphi''(\mu)\,(k_B T)^2 + \mathcal{O}(T^4).\label{eq:Sapp}
	\end{equation}
	
	Let's now write the transport coefficients $\mathcal{L}_{12}$ and $\mathcal{L}_{11}$ in this regime. The non-diagonal transport coefficient $\mathcal{L}_{12}$ in the low temperature limit is written as follow:
	\begin{align}
		\mathcal{L}_{12}\xrightarrow[]{T\to0}\frac{e\pi}{6\hbar} k_B^2T\mathcal{T}'\Big(\frac{eV}{2}\Big)+ \mathcal{O}(T^4)
	\end{align}
	and the diagonal one:
	\begin{align}
		\mathcal{L}_{11}\xrightarrow[]{T\to0} \frac{e^2}{2\pi\hbar}\bigg(\mathcal{T}\Big(\frac{eV}{2}\Big)+\frac{\pi^2}{3}(k_BT)^2\mathcal{T}''\Big(\frac{eV}{2}\Big)+ \mathcal{O}(T^4)\bigg)\label{eq:GLT1}
	\end{align}
	By using these expressions, the final thermopower expression up to the leading order in temperature can be write in the following form:
	\begin{equation}
		Q=\frac{\mathcal{L}_{12}}{\mathcal{L}_{11}}\xrightarrow[]{T\to0} \frac{ \pi ^2 k_B^2 T }{3e}\frac{\mathcal{T}'\big(\frac{eV}{2}\big)}{\mathcal{T}\big(\frac{eV}{2}\big)}=\frac{ \pi ^2 k_B^2 T }{3e}\frac{\partial \log (\mathcal{T}\big(\mu\big)))}{\partial \mu}\bigg|_{\mu=eV/2}.\label{eq:QlowT}
	\end{equation}
	It is worthy noting that, the thermopower expression in low temperature regime shows a similar expression to the Mott formula for the Seebeck coefficient \cite{jonson1980,sivan1986,lunde2005,hou2013}, the difference occurs within transmission coefficient, where the original Mott's formula contain a derivative of logarithm of differential conductance, and the presence of the applied voltage $eV$.
	According to the Ref. \cite{lunde2005}, we are abel to write our thermopower in low temperature regime as logarithmic derivative of the electric conductance, by using the expression \eqref{eq:GLT1} at zero temperature. In this case Eq. \eqref{eq:QlowT} can be written as:
	\begin{equation}
		Q\xrightarrow[]{T\to0} \frac{ \pi ^2 k_B^2 T }{3e}\frac{\partial \log (G\big(\mu,T=0\big))}{\partial \mu}\bigg|_{\mu=eV/2}+ \mathcal{O}(T^3).\label{eq:QlowT2}
	\end{equation}
	Now our expression reduced exactly to the Mott's formula. Also our model reduced to the known one that developed in Refs. \cite{schulz1991,kane1996}.
	
	By using transmission coefficient $\mathcal{T}(\omega)$ given by Eq. \eqref{eq:Tw}, we find:
	\begin{equation}
		Q\xrightarrow[]{T\to0} \frac{ \pi ^2 k_B^2 T }{3e}\frac{4 A^2}{A^2 e V+e^3 V^3},
	\end{equation}
	Mott's formula makes transparant the fact that large $Q$ requires strong particule-hole asymmetry around $\mu$. Transmission changes rapidely with energy around $\mu$, which in the Mott picture enhances $\partial \log G/\partial \mu$, and thus enhances $Q$at low $k_BT$ in accordance with the curves obtaines in figure \ref{fig:QFQH-OCC}. 
	
	Now we examine this expression of thermopower in different regimes of backscattering:
	
	\paragraph*{Strong backscattering regime (SBS):} The SBS regime is verified when the applied voltage is small compared to the backscattering amplitude ($V\ll \tilde{A}$),
	\begin{equation}
		Q\xrightarrow[T\to0]{V\ll \tilde{A}} \frac{ \pi ^2 k_B^2 T }{3e}\frac{4}{eV}\bigg[1-\frac{e^2 V^2}{\tilde{A}^2}\bigg],\label{eq:QSBS}
	\end{equation}

	\paragraph*{Weak backscattering regime (WBS):} In the case of WBS regime is hold when the backscattering amplitude is small compared to  the applied voltage  ($V\gg \tilde{A}$),
	\begin{equation}
		Q\xrightarrow[T\to0]{V\gg \tilde{A}} \frac{ \pi ^2 k_B^2 T }{3e}\frac{4 \tilde{A}^2}{e^3 V^3}\bigg[1-\frac{\tilde{A}^2}{e^2 V^2}\bigg].\label{eq:QWBS}
	\end{equation}
	It is clear that equation (\eqref{eq:QSBS}) gives a maximum value since $V\ll \tilde{A}$, and equation (\eqref{eq:QWBS}) gives a  much smaller value since $V\gg \tilde{A}$. Both cases are obtaines in the Mott/Sommerfeld description in accordance with the linear behavior of $Q$ at low $T$ obtained in the left curves of figure \ref{fig:QFQH-OCC}.
	
	\subsubsection{High temperature limit:}
	
	Next we examine the thermopower in high temperature regime for both systems $k_BT\gg (\hbar\omega\mp eV/2)$, where we start by writting the transport coefficient in the high temperature regime.
	In this regime the non-diagonal transport coefficient $\mathcal{L}_{12}$ is given by:
	\begin{align}
		\mathcal{L}_{12}\xrightarrow[]{T\to\infty}\frac{e^2V \tilde{A}}{32\hbar k_BT^2},
	\end{align}
	and the diagonal one:
	\begin{align}
		\mathcal{L}_{11}\xrightarrow[]{T\to\infty} \frac{e^2}{2\pi\hbar}\bigg(1+\frac{\pi\hbar \tilde{A}}{8k_B T}\Big)
	\end{align}
	After using the above expression we find the final form of thermopower at high temperature limit at leading order in $1/T^2$:
	
	\begin{equation}
		Q\xrightarrow[]{T\to\infty}\frac{\pi V \tilde{A}}{16 k_BT^2}+\mathcal{O}(T^{-3}),\label{eq:QhighT}
	\end{equation}

	\section{Linear Response Regime}\label{sec:LRR}

	Now we aim to relate our thermopower to the Seebeck coefficient in the limit of open circuit. To do that we use the linear limit response which state that, the gradient of applied voltage $V$ and temperature $\Delta T$ are so weak. For that the Fermi-Dirac distribution becomes \cite{crepieux2019}:
	\begin{align}
		f(\hbar\omega-eV/2)&\simeq f_0(\hbar\omega-\varepsilon_F)-\bigg(\frac{eV}{2}+\frac{(\hbar\omega-\varepsilon_F)\Delta T}{2T_0}\bigg)\frac{\partial f_0(\hbar\omega-\varepsilon_F)}{\partial (\hbar\omega)}\label{eq:fLR1}\\
		f(\hbar\omega+eV/2)&\simeq f_0(\hbar\omega-\varepsilon_F)+\bigg(\frac{eV}{2}+\frac{(\hbar\omega-\varepsilon_F)\Delta T}{2T_0}\bigg)\frac{\partial f_0(\hbar\omega-\varepsilon_F)}{\partial (\hbar\omega)}
	\end{align}
	where $\varepsilon_F$ is Fermi energy, $T_0$ is the average temperature of the two reservoirs, and $f_0(\hbar\omega-\varepsilon_F)=(1+exp((\hbar\omega-\varepsilon_F)/(k_BT_0)))$.

	\subsection{ Seebeck coefficient}

	Using these expressions the electric current \eqref{ecurrent} becomes:
	\begin{equation}
		I(V)\simeq -\frac{e}{4\pi\hbar}\int_{-\infty}^{+\infty}\mathcal{T}(\hbar\omega) \bigg(eV+\frac{(\hbar\omega-\varepsilon_F)\Delta T}{T_0}\bigg)\frac{\partial f_0(\hbar\omega-\varepsilon_F)}{\partial (\hbar\omega)}d(\hbar\omega)~.
	\end{equation}
	In this limit, both diagonal and off-diagonal transport coefficients become:
	\begin{align}
		\mathcal{L}_{12}&\simeq-\frac{e}{4\pi}\int_{-\infty}^{+\infty}\mathcal{T}(\hbar\omega) \frac{(\hbar\omega-\varepsilon_F)}{T_0}\frac{\partial f_0(\hbar\omega-\varepsilon_F)}{\partial (\hbar\omega)}d(\hbar\omega),\\
		\mathcal{L}_{11}&\simeq -\frac{e^2}{4\pi}\int_{-\infty}^{+\infty}\mathcal{T}(\hbar\omega) \frac{\partial f_0(\hbar\omega-\varepsilon_F)}{\partial (\hbar\omega)}d(\hbar\omega)~.
	\end{align}
	The above thermopower expression reduces now to Seebeck coefficient \cite{crepieux2019}, and is given by:
	\begin{equation}
		S_{Seebeck}=\frac{\mathcal{L}_{12}}{\mathcal{L}_{11}}=\frac{1}{eT_0}\frac{\int_{-\infty}^{+\infty}\mathcal{T}(\hbar\omega) (\hbar\omega-\varepsilon_F)f_0'(\hbar\omega-\varepsilon_F)d(\hbar\omega)}{\int_{-\infty}^{+\infty}\mathcal{T}(\hbar\omega) f_0'(\hbar\omega-\varepsilon_F)d(\hbar\omega)}~.\label{eq:seeb}
	\end{equation}
	Remarkably, the Peltier coefficient can be concluded from Seebeck coefficient by the Kelvin-Onsager relation $\Pi=T_0 S_{Seebeck}$ \cite{crepieux2019}.
	
	\begin{figure}[!h]
		\begin{center}
			\includegraphics[width=0.32\textwidth]{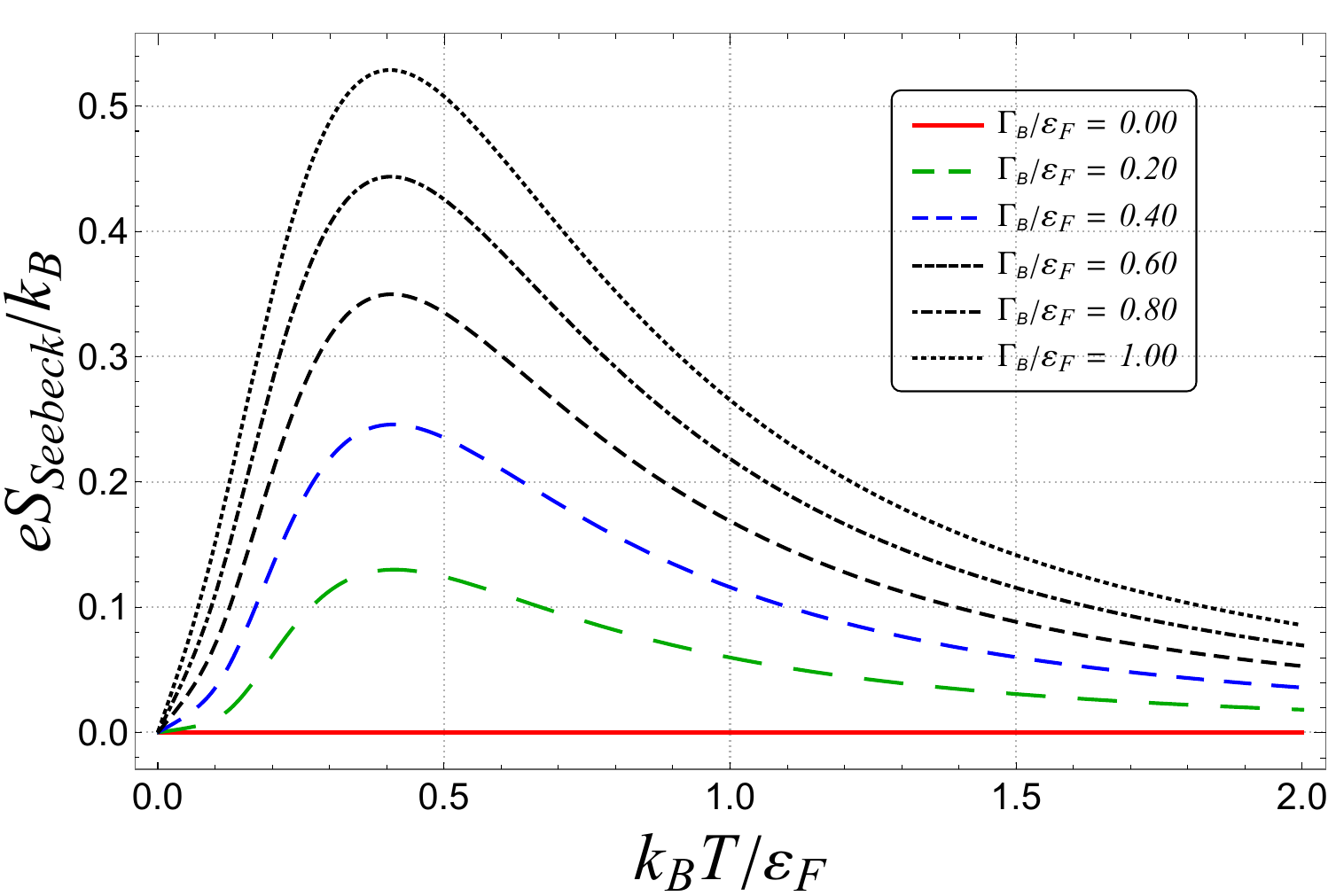}
			\includegraphics[width=0.32\textwidth]{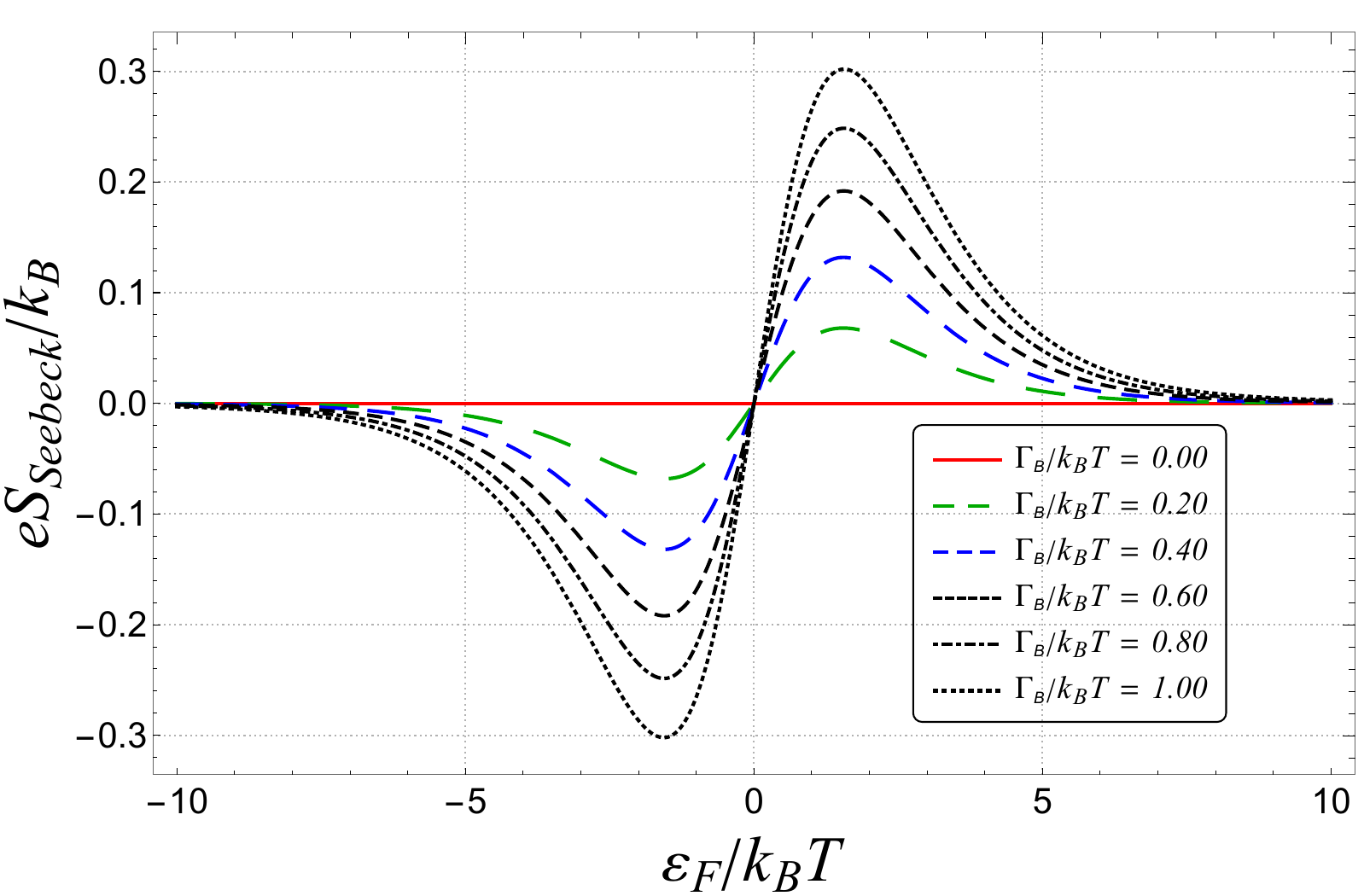}\\
			\includegraphics[width=0.32\textwidth]{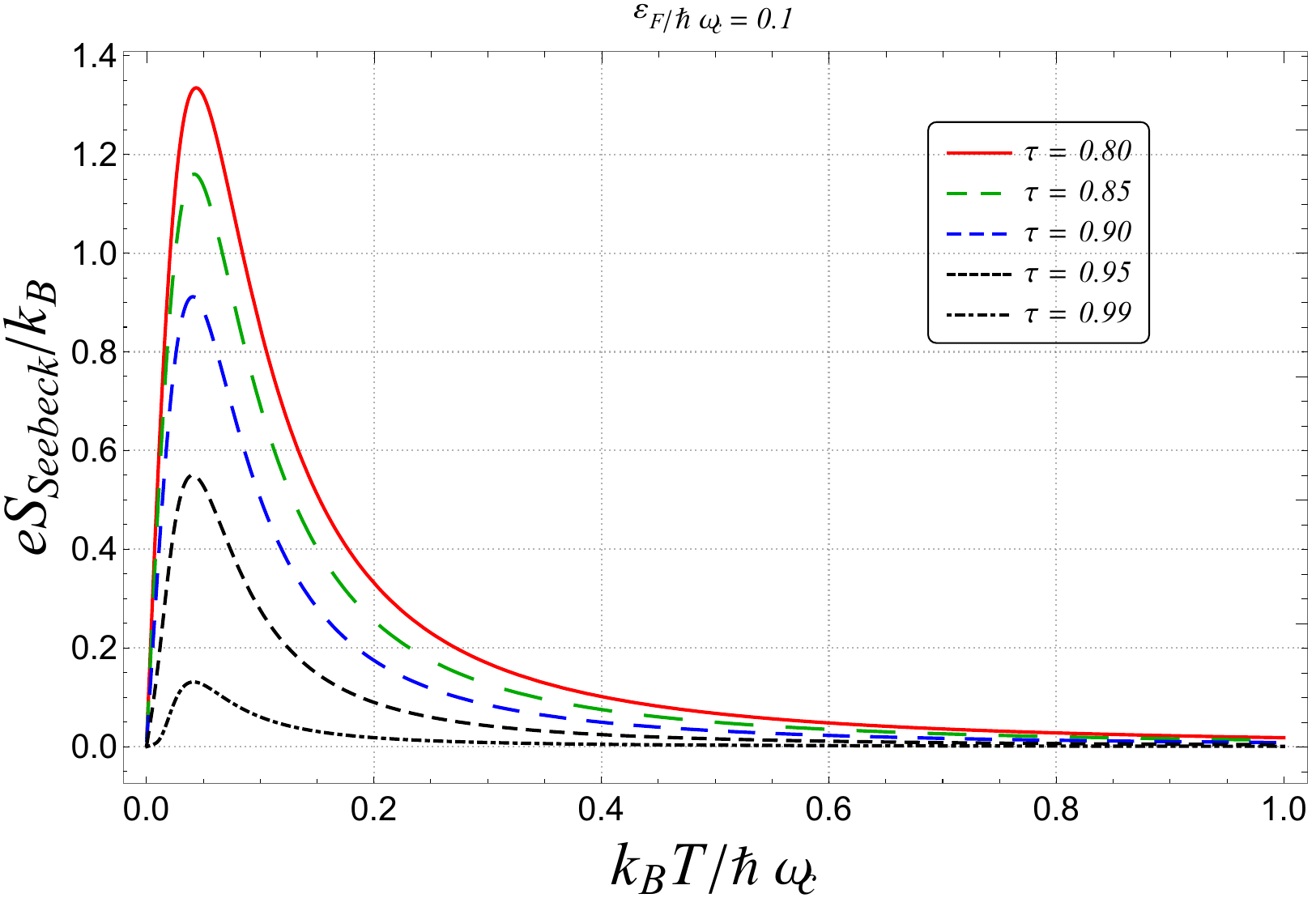}
			\includegraphics[width=0.32\textwidth]{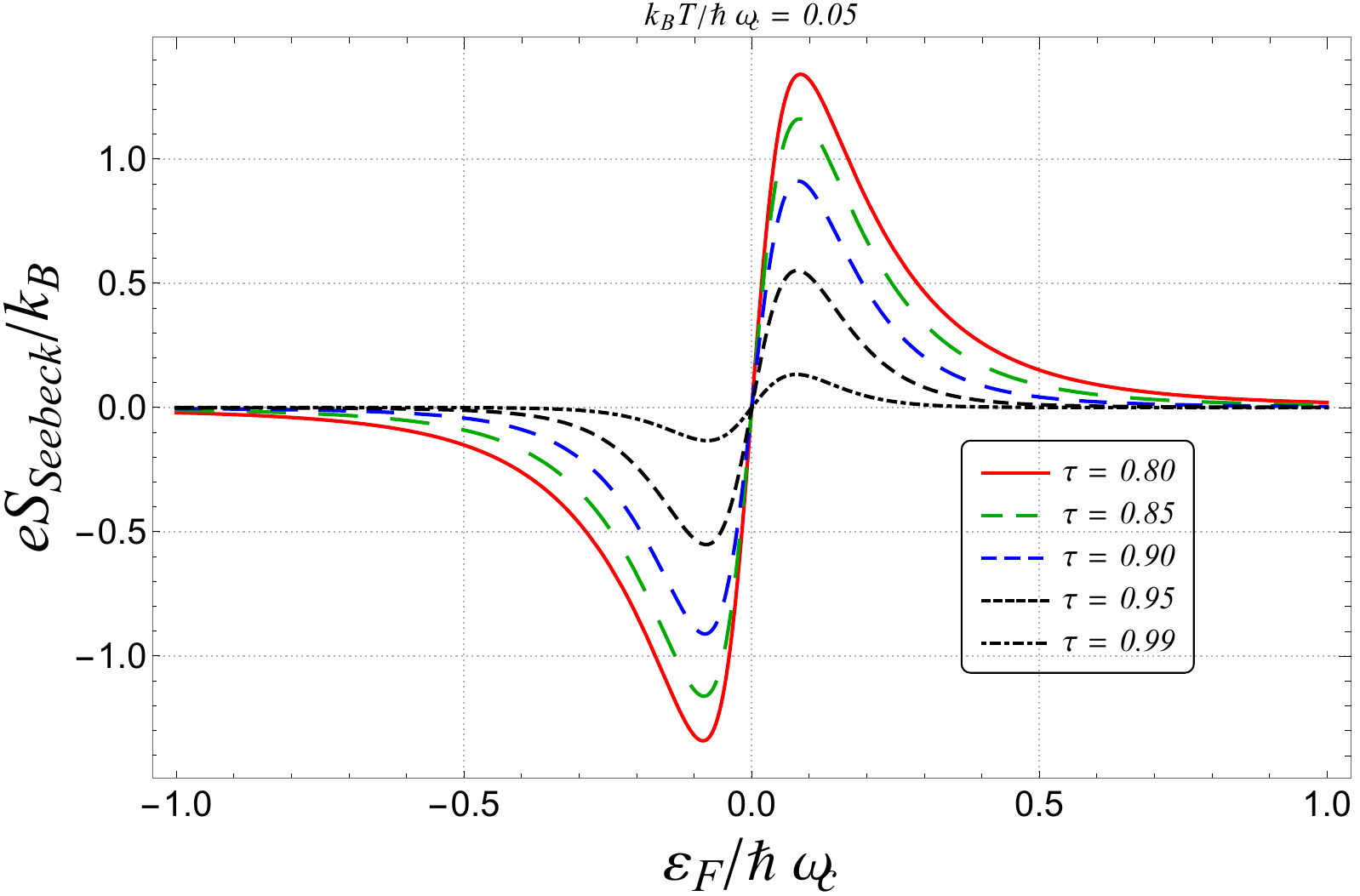}
			\caption{Seebeck coefficient: In first row, versus temperature $k_BT/\varepsilon_F$ in left panel and Fermi energy  $\varepsilon_F/k_BT$ in right panel and for different backscattering amplitude $\Gamma_B/\varepsilon_F$, and $\Gamma_B/k_BT$ for QPC system.  In second row, versus temperature $k_BT/\hbar \omega_c$ in left panel and Fermi energy $\varepsilon_F/\hbar \omega_c$ in right panel and for different effective transmission $\tau$, for OCC system.}\label{fig:SFQH}
		\end{center} 
	\end{figure}
	
	In figure \ref{fig:SFQH}, we plot the Seebeck coefficient $S_{Seebeck}$ as function of temperature $T$ and Fermi energy $\varepsilon_F$, for different values of backscattering amplitude $\Gamma_B$ and the effective transmission $\tau$ for QPC and OCC respectively. At low $T$, the derivative $-\partial f/\partial \omega$ is strongly localized arround $\varepsilon_F$, as a consequence only the states with energy scale close to $\varepsilon_F$ participate to transport. The energy window is verry narrow, there is almost no energy asymmetry, which yields $S_{Seebeck}(t \to 0) \rightarrow 0$. At finite $T$, the thermal window becomes broader $(\hbar \omega -\varepsilon_F) \sim k_B T$, and samples the energy dependence of transmission produced by backscattering in vicinity of the impurity. As a result, $S_{Seebeck}$ increases. The maximum occurs when the asymmetry excitations above and below $\varepsilon_F$ becomes large. Here, the thermal window approaches the characteristic energy $k_BT \sim E_{char}$, and effeciently filters the energy dependant transmission. At heigher T, the thermal window is broad and contributions with opposit energy asymmetries partially compensate, producing decrease of $S_{Seebeck}$. Even in the linear response regime, strong backscattering enhances thermopower since it enhances the energy selectivity of transport channel. In the case of OCC, the same interpretation as the non linear casein terms of effective transmission $\tau$ impact holds. A transport channel with $\tau \approx 1$ reduces the energy dependece of $\mathcal{T}(\omega)$ and consequently reduces $S_{Seebeck}$. On the other hand, for lower $\tau$, the backscattering becomes more pronounced and the transmission becomes more energy selective. Notice that large $S_{Seebeck}$ comes always with small $G$. 
	
	When we look to Fermi energy $\varepsilon_F$, it acts as a tuning parameter for particle-hole asymmetry. Near the symmetric point $\varepsilon_F \approx 0$ we have $S_{Seebeck} \approx 0$. Moving away from this point, the energy asymmetry becomes finite and therefor $\vert S_{Seebeck} \vert > 0$. The curves versus $\varepsilon_F$ show an asymmetric behavior $S_{Seebeck}(-\varepsilon_F)=-S_{Seebeck}(\varepsilon_F)$ indicating that sign reversal comes from underlaying energy symetry of transport problem. As in the non linear regime, in the OCC case, the energy cut-off $\hbar \omega_c$ represents  an upper energy limit of validity of the model. Indeed, for $k_BT \rightarrow \hbar \omega_c$ samples energies close to the cut-off.
	
	The linear response is obtained in the zero bias limit $S_{Seebeck}(T)=\lim_{V \to 0} Q(V,T)$. In the linear response regime, $S_{Seebeck}(T)$ probes the local energy asymetry around $\varepsilon_F$. The obtained curves in figure \ref{fig:SFQH} explain local energy dependence of transmission at Fermi energy. However, non linear thermopower $Q$ is sensitive to the transmission over an extanded energy inteval. The analysis of curves in figures \ref{fig:QFQH-OCC} and \ref{fig:SFQH} shows that $Q$ is much larger then $S_{Seebeck}$. This means that a finit voltage allows the system to sample a much larger energy interval, that can produce a large energy imbalance and therfore large non linear thermopower $Q$.   
	
	Now we investigate the Seebeck coefficient in different temperature regimes, as the thermopower above.

	\subsubsection{Low temperature limit}
	By applying the same approximation used in the non linear  thermopower $Q$ \eqref{eq:Sapp}, the Seebeck coefficient at low temperature regime ($k_B T<<(\hbar\omega-\varepsilon_F)$) becomes:
	\begin{equation}
		S_{Seebeck}\xrightarrow[]{T\to0} \frac{ \pi ^2 k_B^2 T }{3e}\frac{\mathcal{T}'\big(\varepsilon_F\big)}{\mathcal{T}\big(\varepsilon_F\big)}\simeq \frac{ \pi ^2 k_B^2 T }{3e} \frac{\partial \log \mathcal{T}\big(\varepsilon\big)}{\partial \varepsilon}\bigg|_{\varepsilon=\varepsilon_F},\label{eq:SClowT}
	\end{equation}
	which is exactly similar to the Mott formula \cite{jonson1980,sivan1986,lunde2005,hou2013}, here the difference remains in presence of the transmission coefficient $\mathcal{T}(\varepsilon)$, instead of differential conductance. Its worthy noting that, the thermopower \eqref{eq:QlowT} and Seebeck coefficient \eqref{eq:SClowT} coincide in the low temperature regime in the low voltage limit (around Fermi energy) in which $eV/2\to \varepsilon_F$.\\
	The Mott's formula shows clearly that $S_{Seebeck}$ is sensitive to energy dependence of $\mathcal{T}(\omega)$. Thus, stronger backscattering produces a stronger energy dependence of $\mathcal{T}(\omega)$ which increases $\partial \log \mathcal{T}/\partial \varepsilon $. The enhancement of $\tilde{A}$ generates the enhancement of $S_{Seebeck}$. \\
	The fact that $S_{Seebeck} \propto T$ in the Mott's formula under a Sommerfeld expension, provide to us that transmission is regular at $\varepsilon=\varepsilon_F$. This result provides a consistency check for numerical results, since the curves in figure \ref{fig:SFQH} show a linear low temperature regime. 
	
	\subsubsection{High temperature limit}
	In the high temperature limit the Seebeck coefficient becomes:
	\begin{equation}
		S_{Seebeck}\xrightarrow[]{T\to\infty}\frac{\pi \varepsilon_F \tilde{A}}{8e k_BT^2}+\mathcal{O}(T^{-3}),\label{seebhigh}
	\end{equation}
	this expression shows again a similarity with the thermopower with difference in the applied voltage, if $eV/2\to \varepsilon_F$ the thermopower expression \eqref{eq:QhighT} reduced directly to the Seebeck coefficient (\eqref{seebhigh}) in the high energy limit.

	\subsection{Entropy variation associated to thermoelectric particle transport }

	The thermopower of the system is related to the thermodynamical  entropy per carrier through the Kelvin formula for thermopower \cite{peterson2010}, and is defined by \cite{barlas2012,strunk2021,cortes2023,goupil2011}:
	\begin{equation}
		s=\frac{\partial S}{\partial N}=eS_{Seebeck},\label{eq:epp1}
	\end{equation}
	where $e$ is the charge of the electron that carries the interaction. In order to verify the applicability of this relation in our system, we compare the Seebeck coefficient and the entropy per carrier, depending on density of state $D(\varepsilon)$, and given by \cite{tsaran2017,galperin2018,grassano2018,cortes2023}:
	\begin{equation}
		s=\frac{1}{T_0}\frac{\int_{-\infty}^{+\infty}D(\hbar\omega) (\hbar\omega-\varepsilon_F)f_0(\hbar\omega-\varepsilon_F)'d(\hbar\omega)}{\int_{-\infty}^{+\infty}D(\hbar\omega) f_0(\hbar\omega-\varepsilon_F)'d(\hbar\omega)},\label{eq:epp2}
	\end{equation}
	where $D(\omega)$ represents the density of stats. In TLL with single impurity and interaction parameter $K=1/2$ the density of states reads  \cite{kane1992,vondelft1998,zamoum2014}:
	\begin{equation}
		D(\omega) \propto \hbar |\omega|,\label{eq:DOS1}
	\end{equation}
	
	\begin{figure}[!h]
		\begin{center}
			\includegraphics[width=0.32\textwidth]{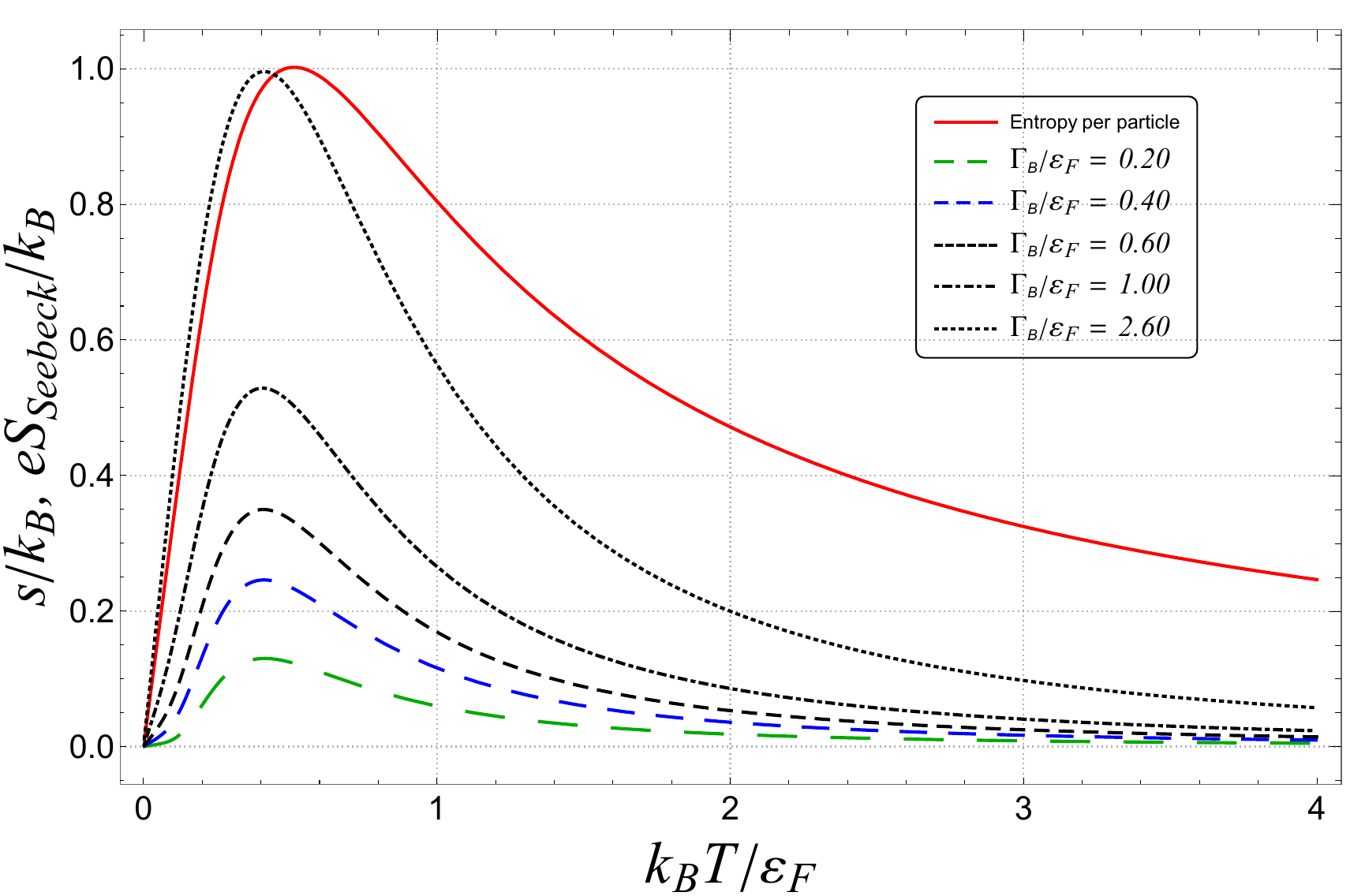}
			\includegraphics[width=0.32\textwidth]{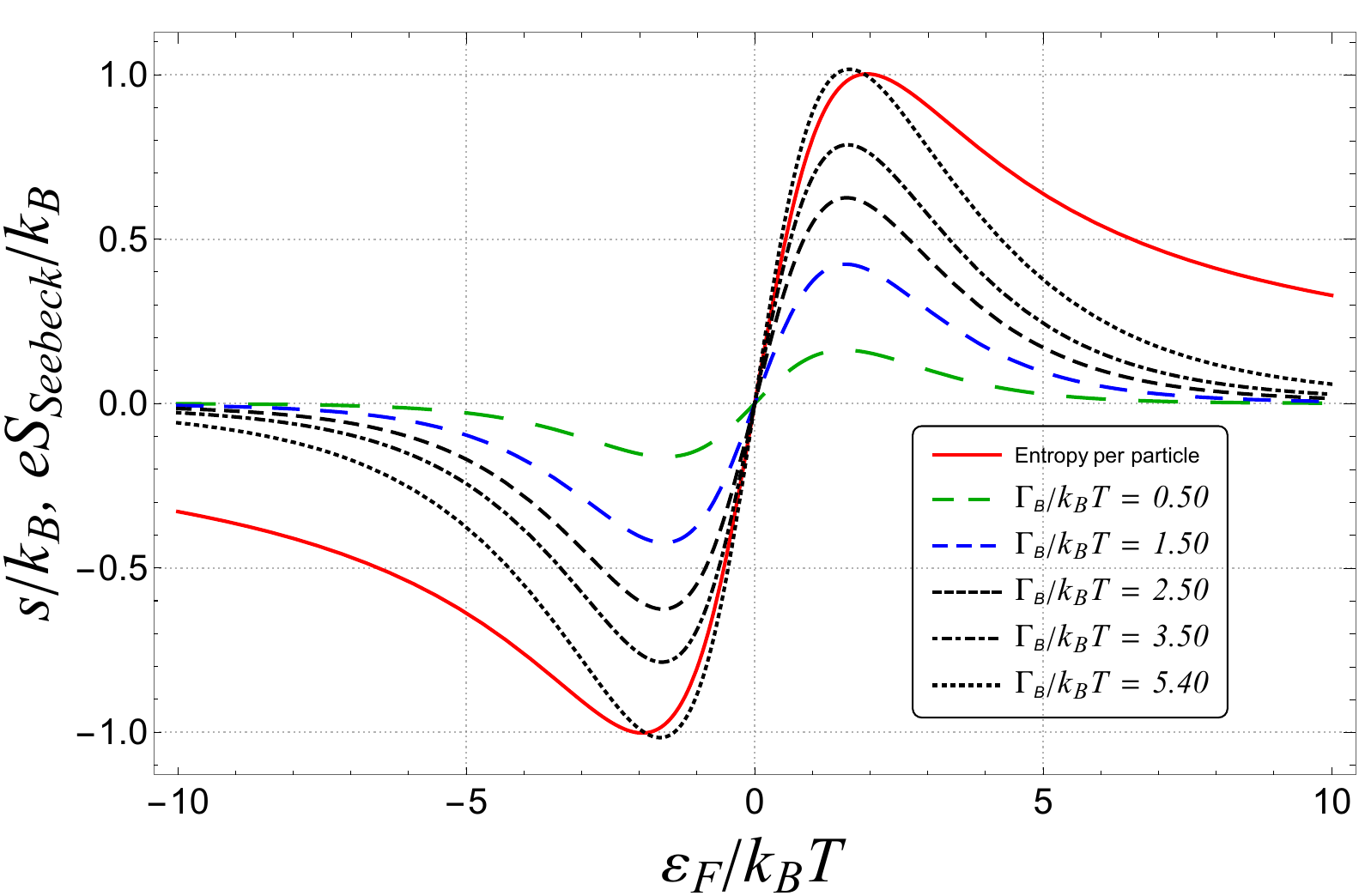}\\
			\includegraphics[width=0.32\textwidth]{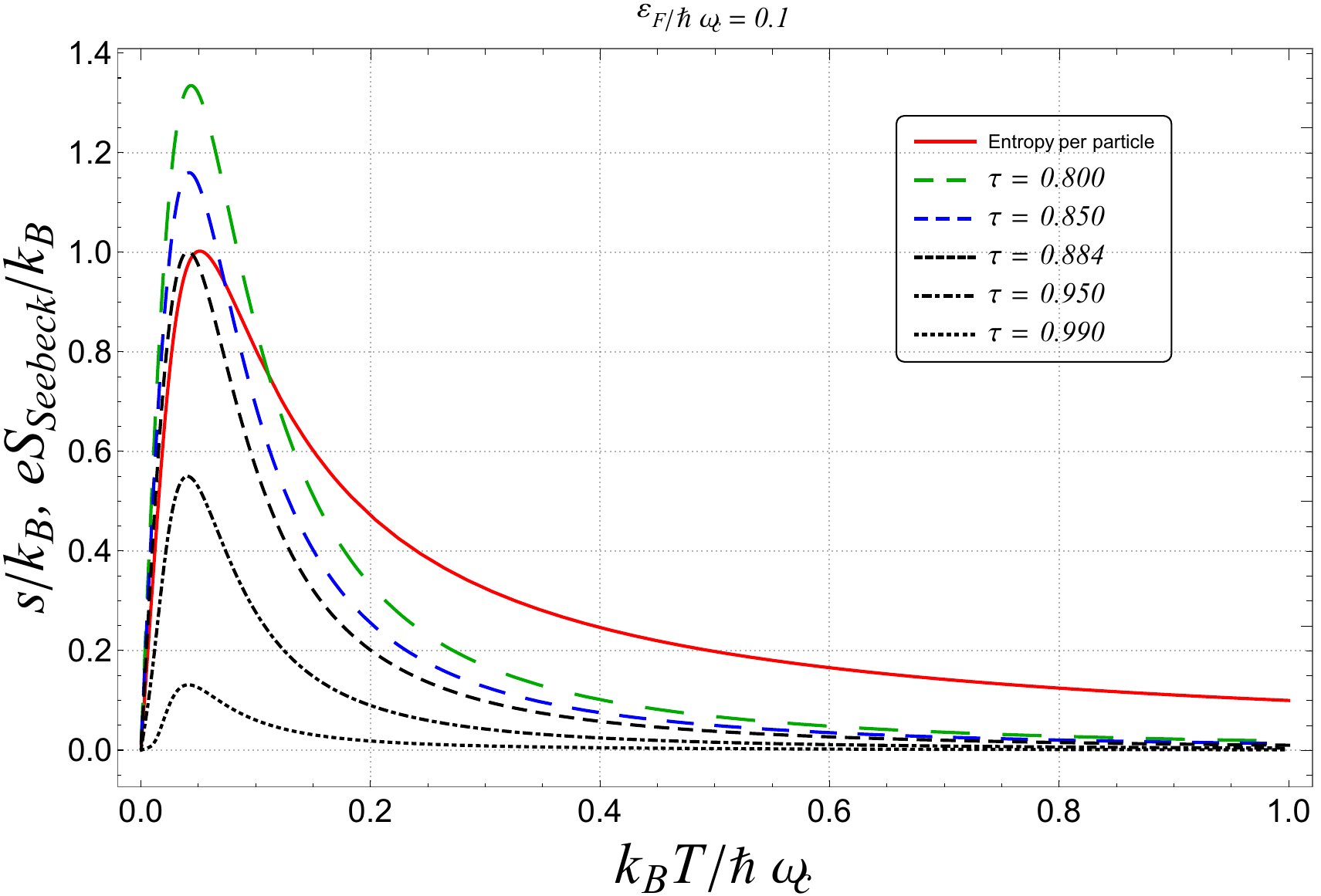}
			\includegraphics[width=0.32\textwidth]{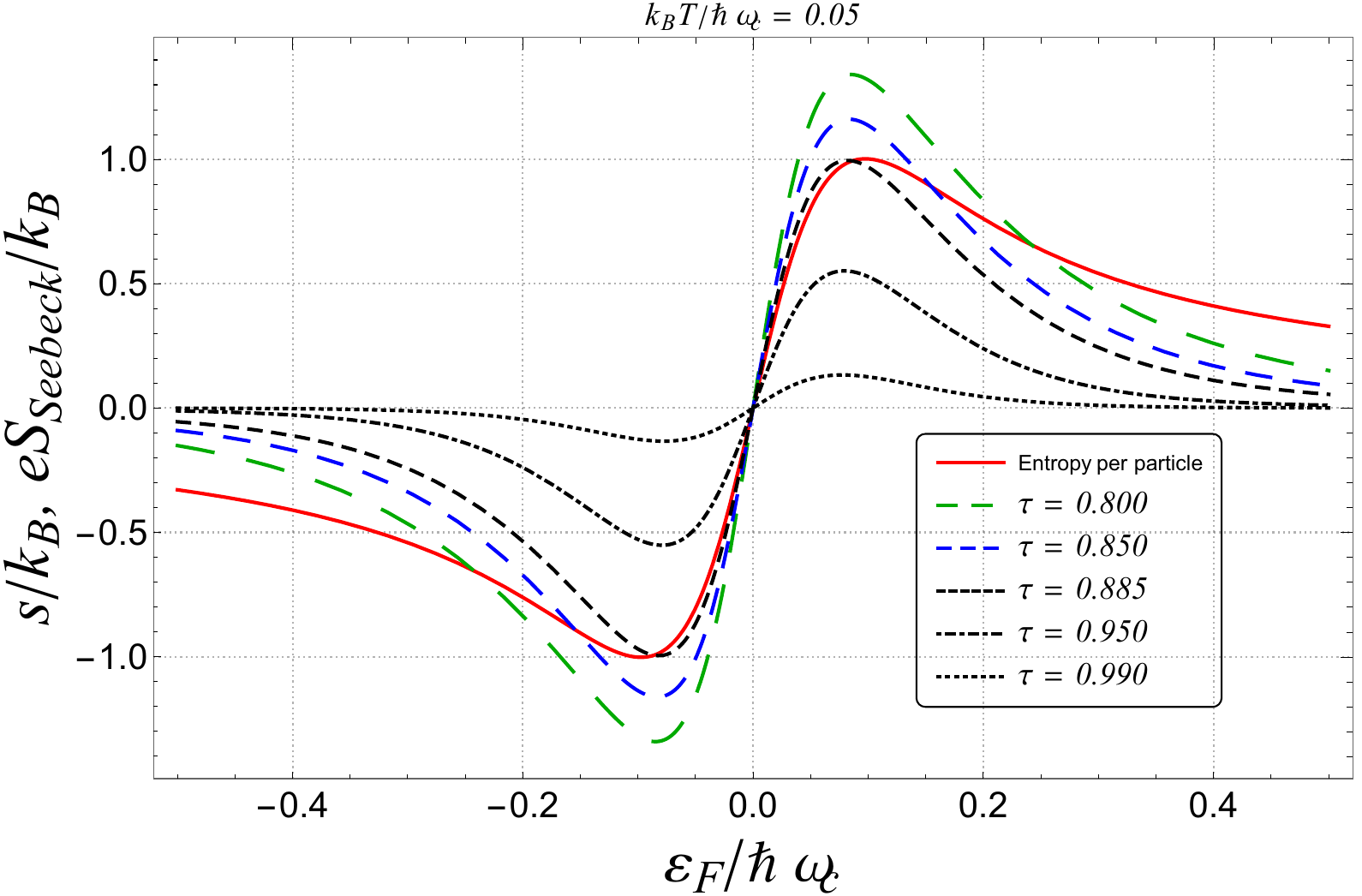}		
			\caption{Seebeck coefficient comparing with Entropy per particle for QPC system versus temperature $k_BT/\varepsilon_F$ in left panel and Fermi energy $\varepsilon_F/k_BT$ in right panel and for different parameters $\Gamma_B/\varepsilon_F$, and $\Gamma_B/k_BT$ in first row. For OCC system versus temperature $k_BT/\hbar \omega_c$ in left panel and applied voltage $eV/\hbar \omega_c$ in right panel and for different effective transmission $\tau$.}\label{fig:ESFQH}
		\end{center} 
	\end{figure}
	In Fig. \ref{fig:ESFQH}, we compare Seebeck coefficient to entropy per particle. The figures are plotted as function of temperature (left panels), and Fermi energy (right panels) for both systems. The graphics show that the two quantities are comparable at low temperature and for specific scale of Fermi energy.

	In order to obtain the total entropy contribution of thermoelectric particle transport, we choose to work within homogeneity postulate \cite{strunk2021}, we use local densities rather than extensive quantities. The particle density is given by \cite{galperin2018}:
	\begin{equation}
		\mathcal{N}= \int_{0}^{+\infty}D(\omega)f(\hbar\omega-\mu)d(\hbar\omega)\,,
	\end{equation}
	where $\mathcal{N}=N/L$, $N$ represents the total number of particles and $L$ is the size of TLL. Using the above density of state given by \eqref{eq:DOS1} and taking into account the dimension of the constants, the particle density take the following form:
	\begin{equation}
		\mathcal{N}=-\frac{1}{\pi\hbar v_F\varepsilon_F}(k_BT)^2 \, Li_2(-e^{\mu/k_BT}) \label{eq:Ntotal}
	\end{equation}
	where $Li_n(z)$ is polylogarithm function.
	Now from the definition of the entropy per carrier given by \eqref{eq:epp1}, we can write using Maxwell formula \cite{peterson2010}:
	\begin{align}
		\frac{\partial \mathcal{S}}{\partial \mathcal{N}}=\frac{\big(\frac{\partial \mathcal{S}}{\partial \mu}\big)}{\big(\frac{\partial \mathcal{N}}{\partial\mu}\big)}=s,\qquad\Longrightarrow \qquad\frac{\partial \mathcal{S}}{\partial \mu}=s\,\bigg(\frac{\partial \mathcal{N}}{\partial \mu}\bigg),
	\end{align}
	Now by integrating the  Kelvin formula for Seebeck coefficient over the variable $\mu=eV/2$, we can obtain the expression of the density of entropy variation related to thermoelectric transport:
	\begin{equation}
		\Delta \mathcal{S}=	\mathcal{S}-\mathcal{S}_0=s\,\int_{\varepsilon_F-eV/2}^{\varepsilon_F+eV/2}\bigg(\frac{\partial N}{\partial \mu}\bigg)d\mu=s\,\bigg(N\bigg(\varepsilon_F+eV/2\bigg)-N\bigg(\varepsilon_F-eV/2\bigg)\bigg),
	\end{equation}
	Using the expressions of entropy per carrier and particle density given by Eqs. \eqref{eq:seeb} and \eqref{eq:Ntotal} respectively, density of entropy variation related to thermoelectric transport become:
	\begin{align}
		\Delta \mathcal{S}&\simeq-\frac{k_B^2T_0}{\pi v_F\hbar \varepsilon_F}\frac{\int_{-\infty}^{+\infty}\mathcal{T}(\hbar\omega) (\hbar\omega-\varepsilon_F)f_0(\hbar\omega-\varepsilon_F)'d(\hbar\omega)}{\int_{-\infty}^{+\infty}\mathcal{T}(\hbar\omega) f_0(\hbar\omega-\varepsilon_F)'d(\hbar\omega)}\bigg(Li_2(-e^{(\varepsilon_F+eV/2)/k_BT})-Li_2(-e^{(\varepsilon_F-eV/2)/k_BT})\bigg),\label{eq:eT1}
	\end{align}
	This is the final expression of the contribution of transmitted particles to the entropy variation for TLL system with impurity and interaction parameter $K=1/2$. 
	\begin{figure}[!h]
		\begin{center}
			\includegraphics[width=0.32\textwidth]{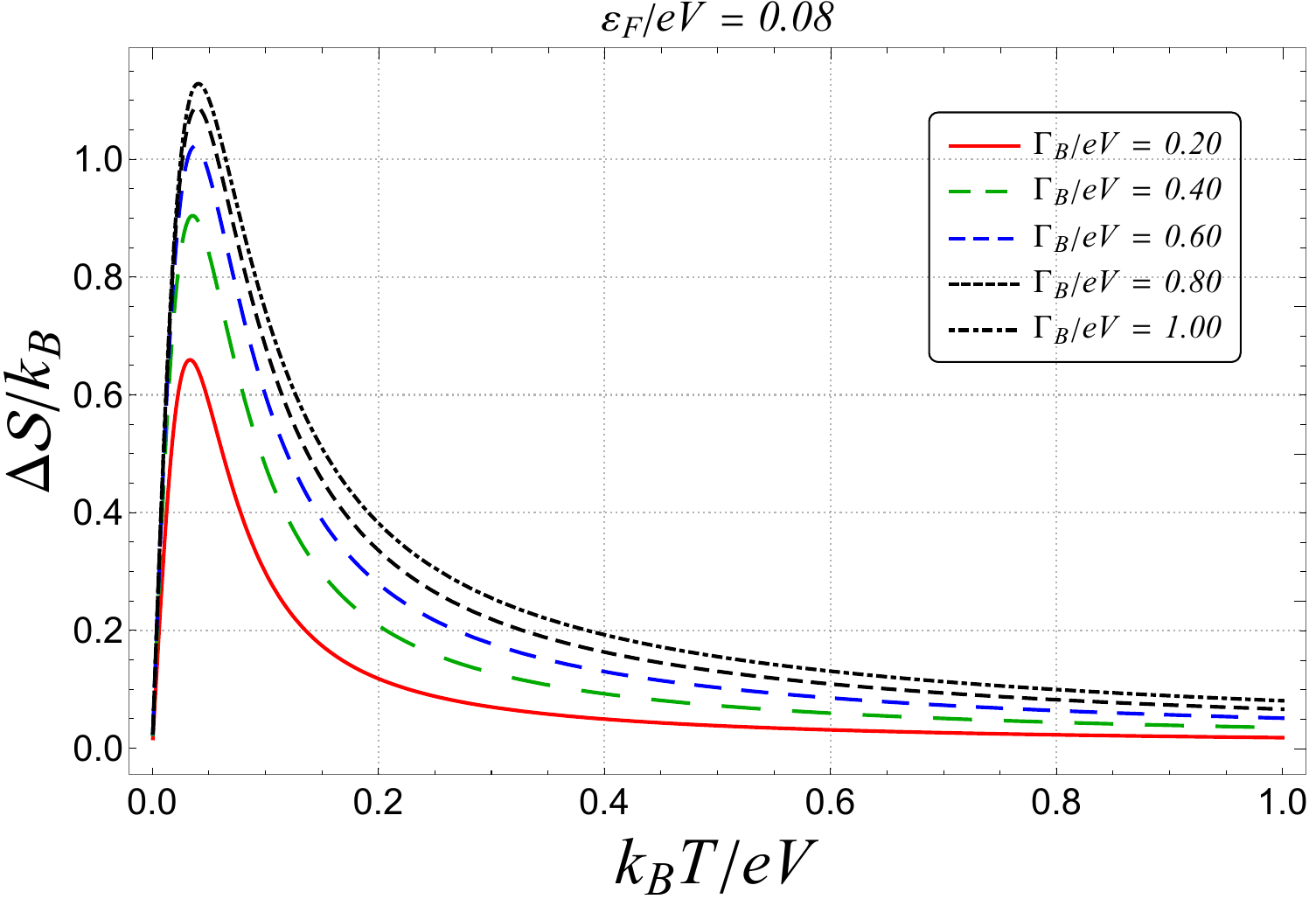}
			\includegraphics[width=0.32\textwidth]{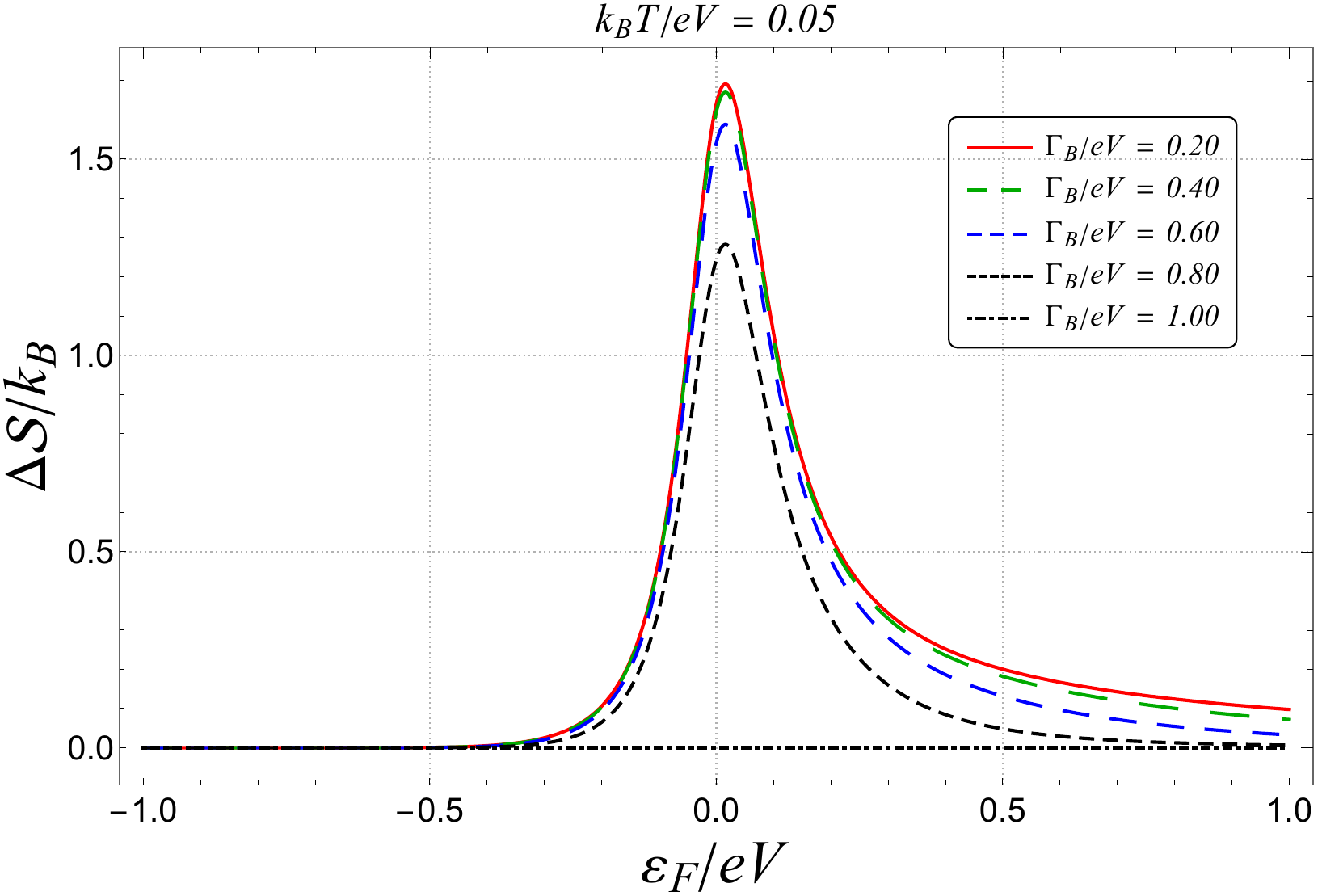}\\
			\includegraphics[width=0.32\textwidth]{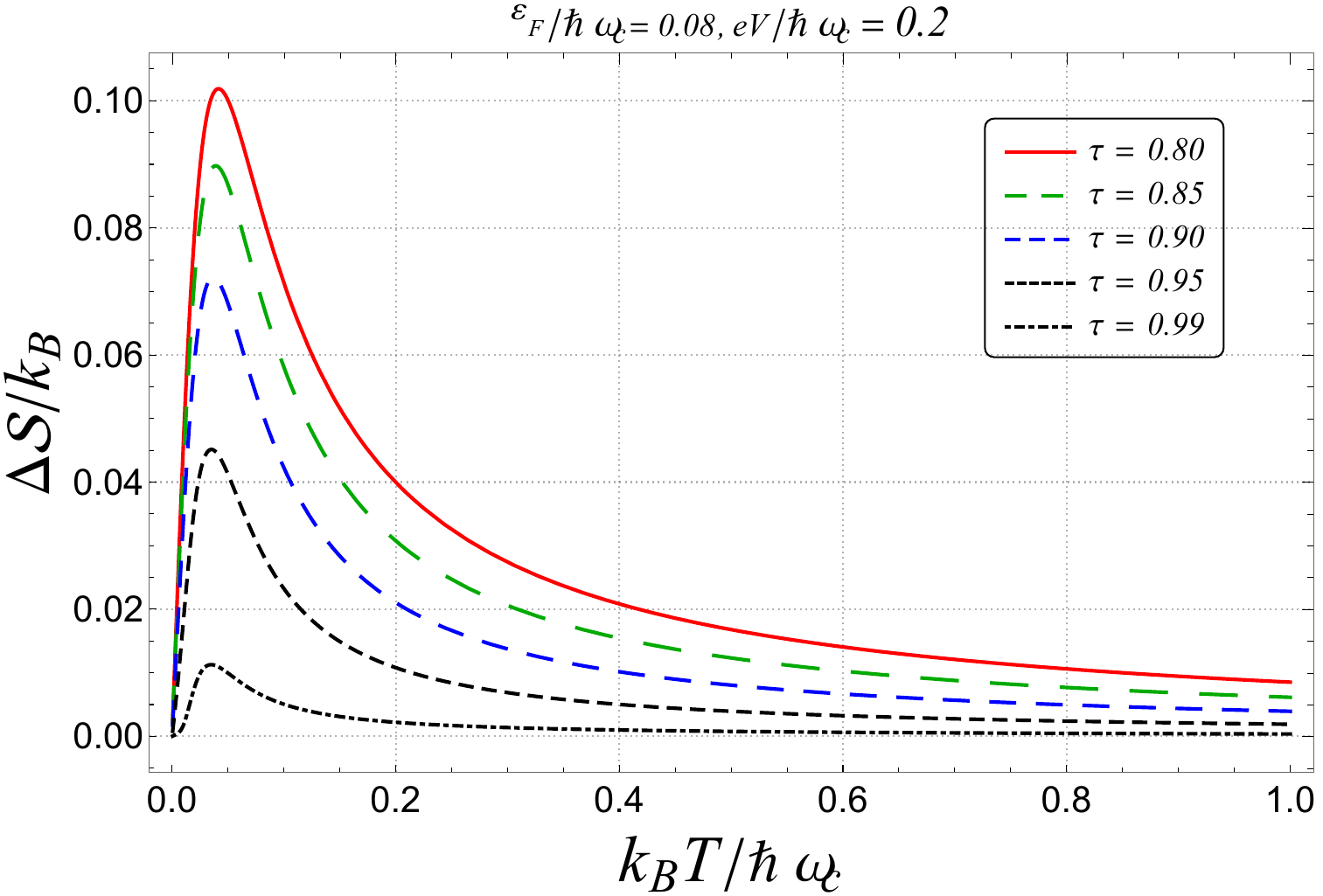}
			\includegraphics[width=0.32\textwidth]{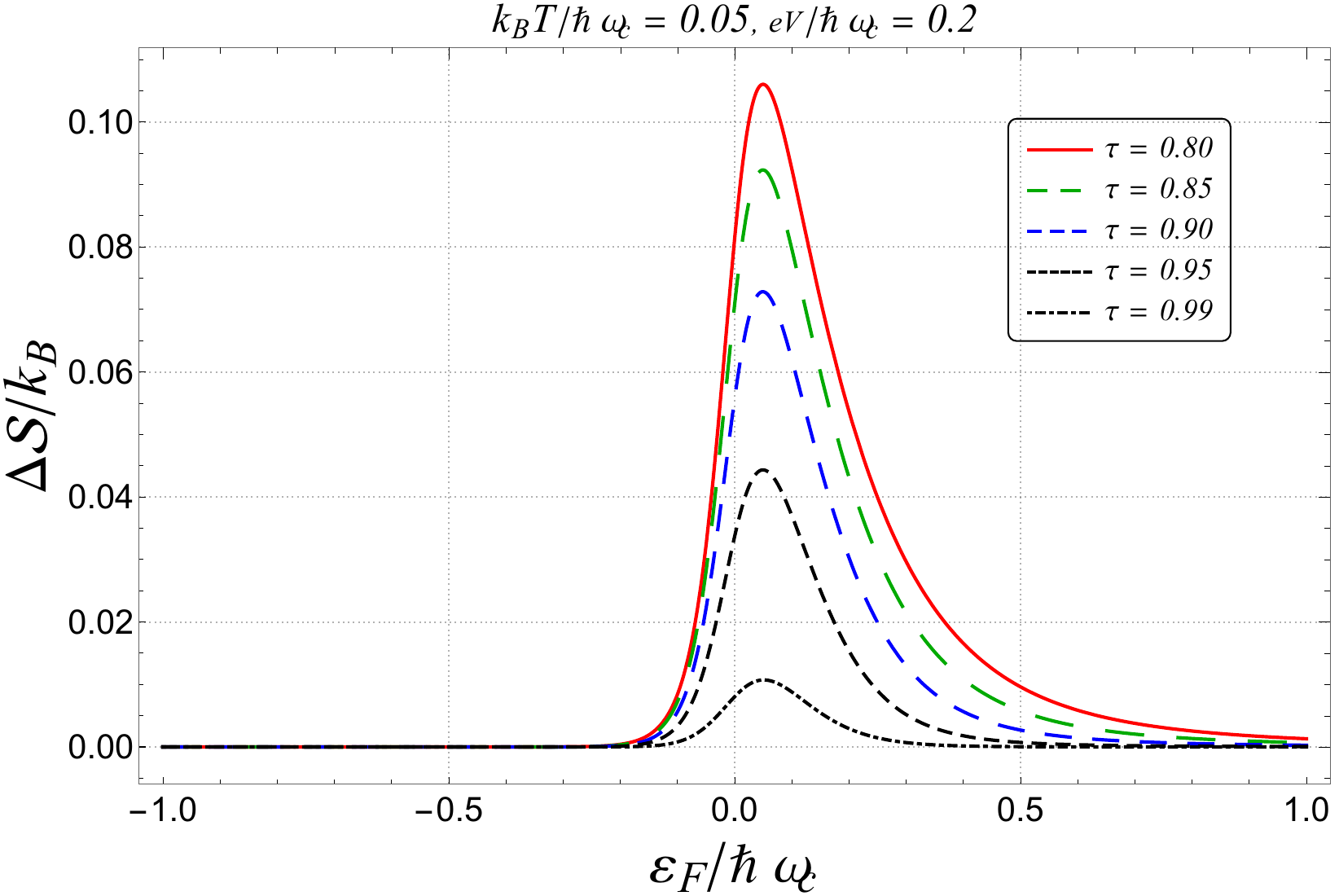}
			\caption{Difference of entropy density $\Delta \mathcal{S}/(k_B)$ versus temperature $k_BT/eV$ in left panel and Fermi energy $\varepsilon_F/eV$ in right panel and for different parameters $\Gamma_B/eV$ for QPC system in first row. In the second row: versus $k_BT/\hbar \omega_c$ in left pannel, ans $\varepsilon_F/\hbar \omega_c$ in the right pannel for different effective transmission $\tau$ for OCC system.}\label{fig:EFQH}
		\end{center} 
	\end{figure}
	
	The entropy variation $\Delta	\mathcal{S}$ is associated with thermoelectric particle transport. Therefor, a large $S_{Seebeck}$ corresponds to a large entropy carried per transmitted particle. 
	
	In the left panels of figure \ref{fig:EFQH} the density of entropy variation related to transport $\Delta\mathcal{S}$ is plotted versus temperature for QPC (upper graphic) and OCC (lower graphic). At $T \rightarrow 0$ the number of thermally accessible excitations is very small, we have $\Delta	\mathcal{S} \rightarrow 0$. At increasing $T$, more states arround $\varepsilon_F$ become thermally accessible, the entropy variation associated with particle transfer increases. A maximum of $\Delta	\mathcal{S}$ occurs when the thermal window becomes comparable to energy scale over which the entropy is strongly energy dependent. The thermally accessible states probe the full energy structure of backscattering region, which produce the largest $\Delta	\mathcal{S}$. At higher energy, the thermal window becomes very broad, contributions from different energies begin to average out, so $\Delta	\mathcal{S}$ decreases. 
	
	For QPC, stronger backscattering produces stronger energy dependence of $\mathcal{T}(\omega)$. As consequence, enhanced $S_{Seebeck}$ produces enhanced $\Delta	\mathcal{S}$. For OCC, smaller $\tau$ produces larger $\Delta	\mathcal{S}$, since $\tau$ corresponds to stronger effective backscattering, which is consistent with a stronger energy filtering. Thus, the backscattering region modifies the energy selectivity of transport, which changes the entropy variation associated with transmitted particle. The backscattering region can be viewed as a source of additional thermodynamic structure. 
	
	In the right side of figure \ref{fig:EFQH}, the dependence of $\Delta	\mathcal{S}$ on Fermi energy $\varepsilon_F$ is plotted. This dependence indicates that transport can be controlled by $\varepsilon_F$ tuning. The maximum corresponds to a particular range of $\varepsilon_F$ when energy asymmetry is strongest. Notice that this contribution to entropy is not the impurity entropy $S_{imp}$ \cite{kattel2026}, which represents the impurity contribution to equilibrium entropy.
	
	Now we can examine this expression at low temperature regime using Sommerfeld approximation \eqref{eq:Sapp}, after some algebra we find:
	\begin{equation}
		\Delta	\mathcal{S}\xrightarrow[]{T\to0} \frac{\pi eVL k_B^2 T_0}{3 v_F \hbar}\frac{\partial\ln(\mathcal{T}(\varepsilon))}{\partial \varepsilon}\bigg|_{\varepsilon=\varepsilon_F}.\label{eq:EntropyLowT}
	\end{equation}
	This relation can be interpreted as Mott's-like formula for entropy variation related to thermoelectric transport.
	
	%--------------------------------------------------------------------
	\subsection{Thermoelectric transport heat capacity}
	%--------------------------------------------------------------------

	The second important materiel property is the thermoelectric transport heat capacity density $\mathcal{C}$, which measures the temperature sensitivity of the entropy variation associated with transport. In the same thermodynamic framework we calculate the density of the this contribution to heat capacity. Schematically, we can write $\mathcal{C}_{total}=\mathcal{C}_{bulk}+\mathcal{C}$. Consequentially, $\mathcal{C}$ is seen as a correction to total heat capacity.

	Using the standard definition of heat capacity and equation Eq. \eqref{eq:eT1}, we write:
	\begin{align}
		\mathcal{C}=T_0\frac{\partial \Delta S}{\partial T_0},\label{eq:TC1}
	\end{align}
	\begin{figure}[!h]
		\begin{center}
			\includegraphics[width=0.32\textwidth]{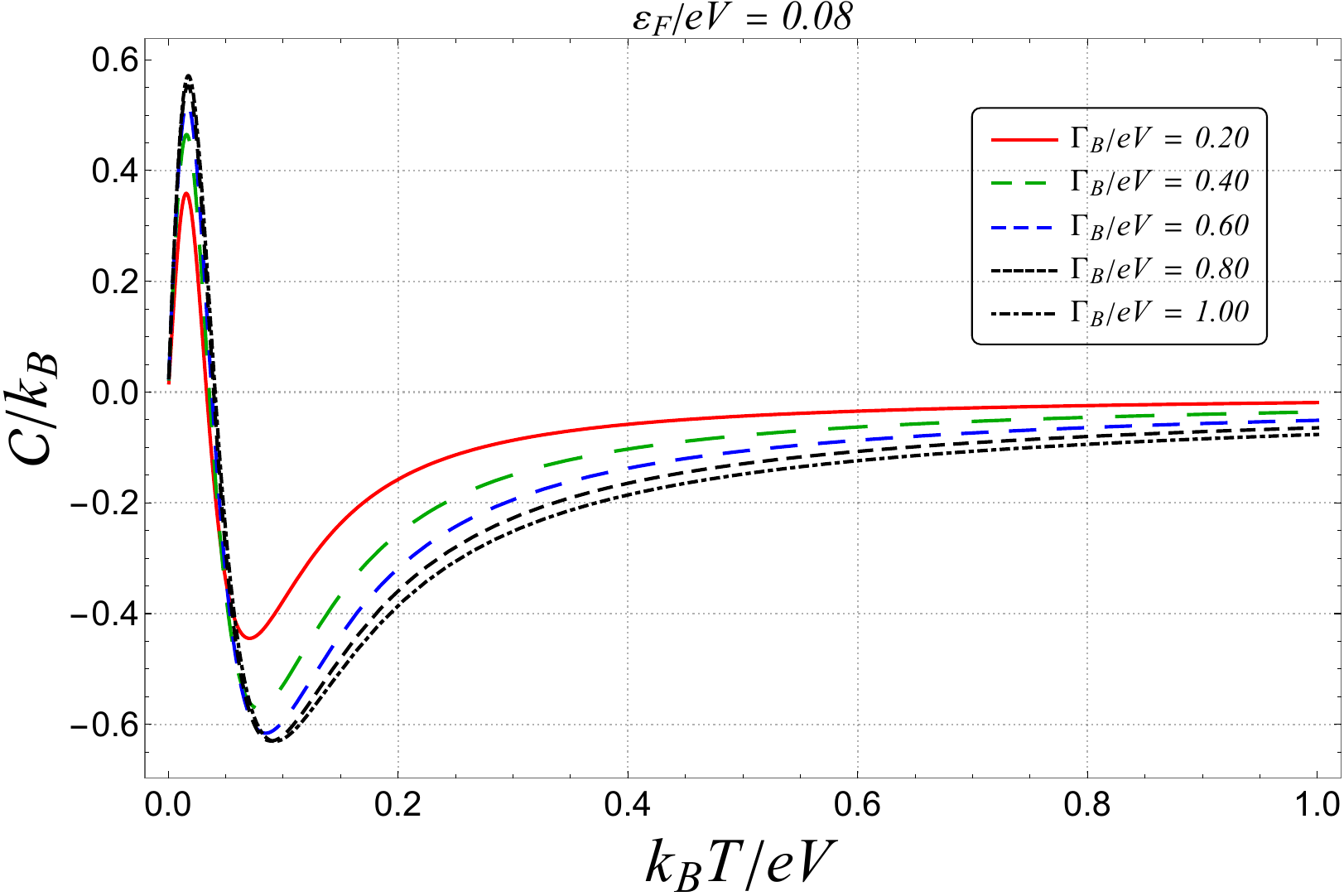}
			\includegraphics[width=0.32\textwidth]{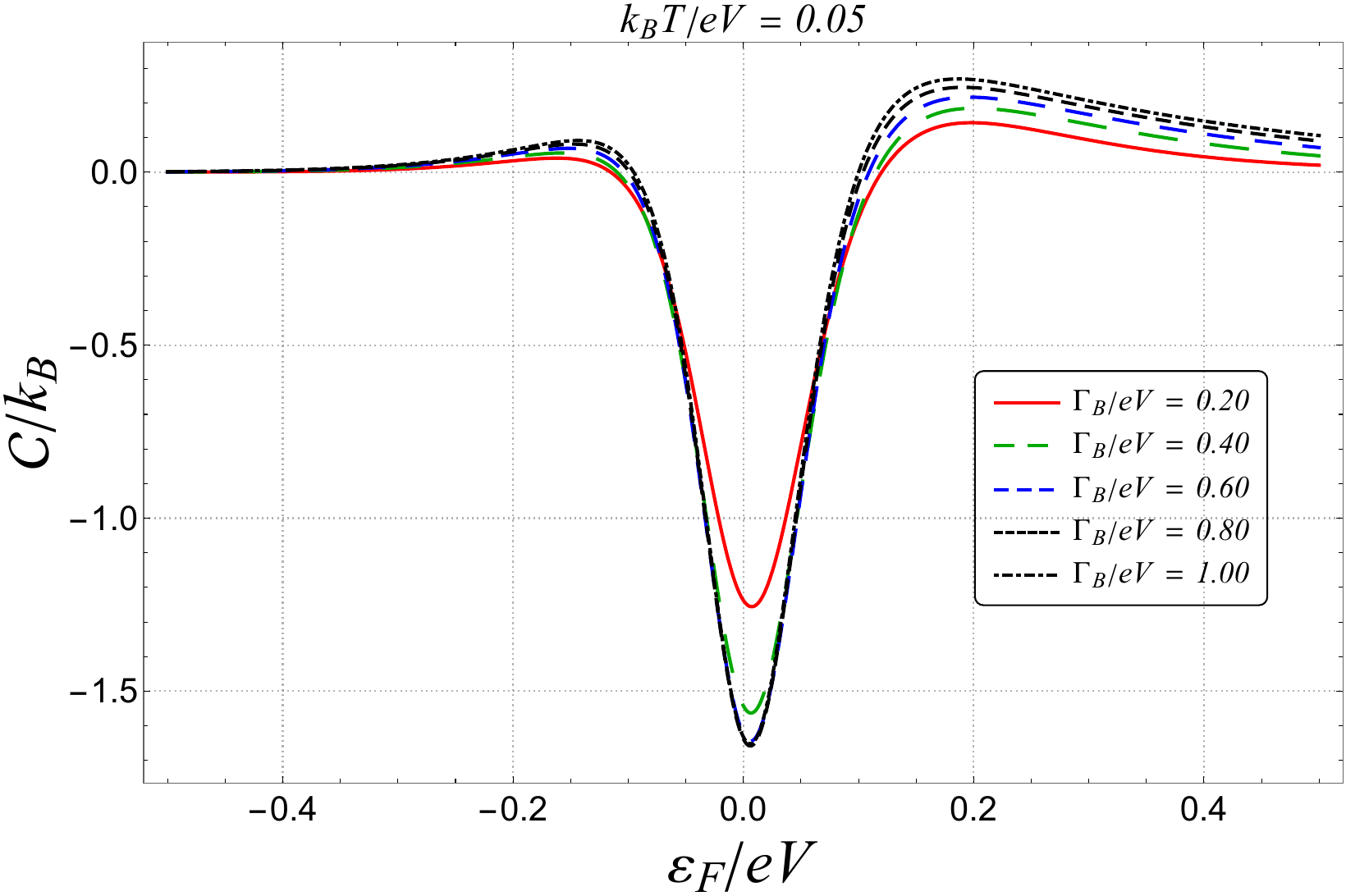}\\
			\includegraphics[width=0.32\textwidth]{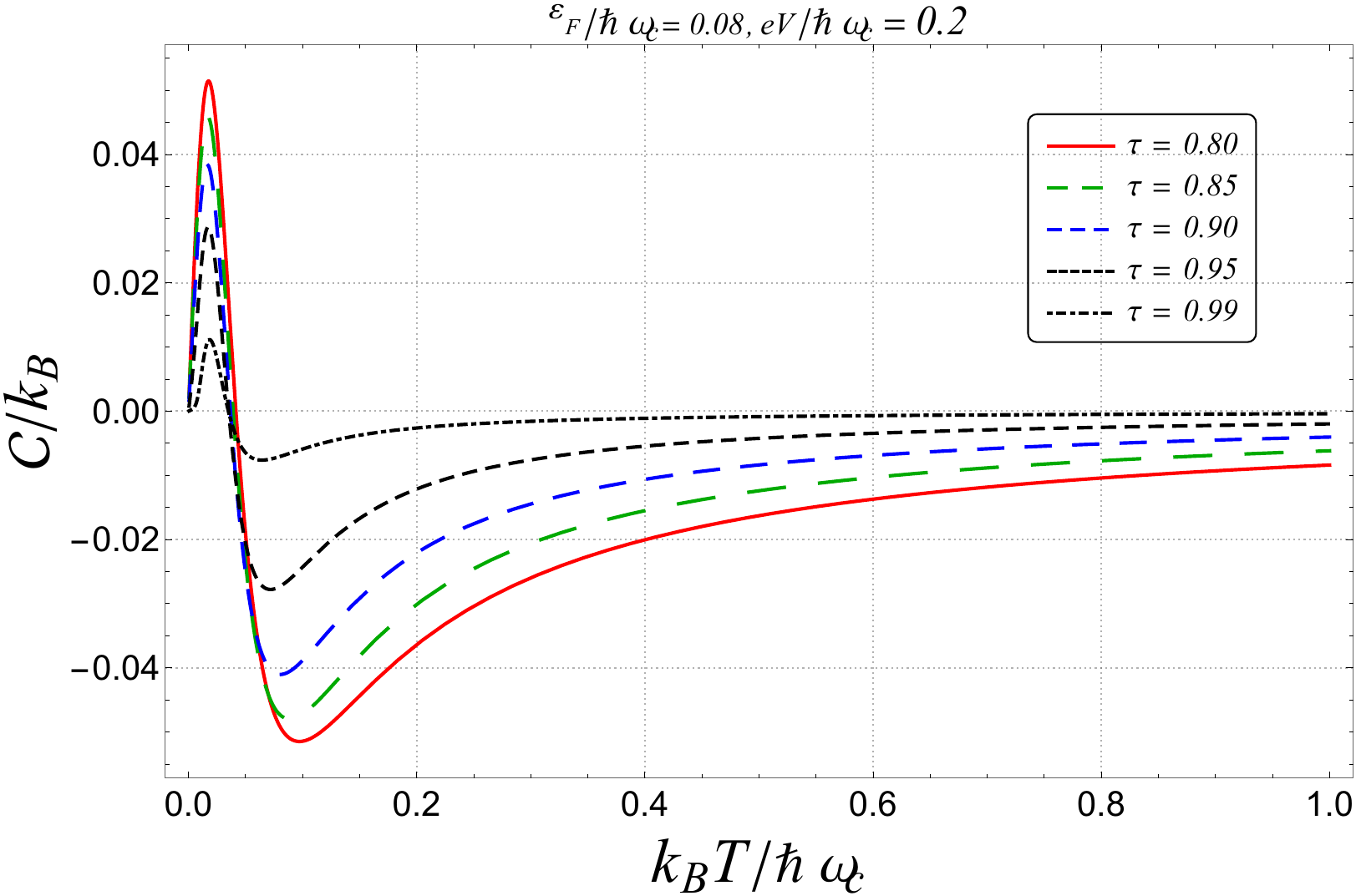}
			\includegraphics[width=0.32\textwidth]{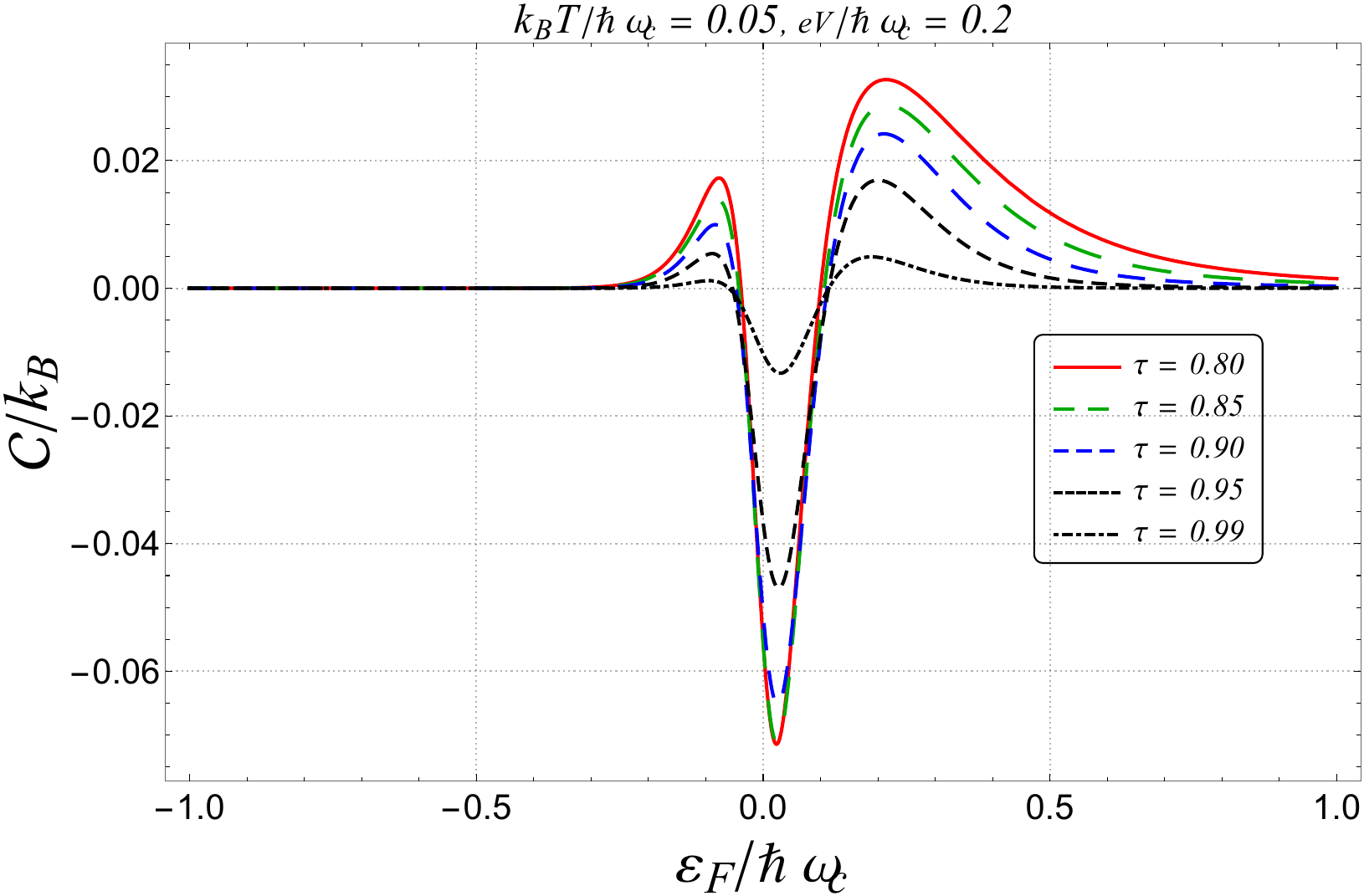}
			\caption{thermoelectric transport heat capacity density correction $\mathcal{C}/(k_B)$ versus temperature $k_BT/eV$ in left panel and Fermi energy $\varepsilon_F/eV$ in right panel and for different parameters $\Gamma_B/eV$, for QPC system in first row. In second row: versus $k_BT/\hbar \omega_c$ in left pannel, and versus $\varepsilon_F/\hbar \omega_c$ in the right pannel for different effective transmission $\tau$ for OCC system.}\label{fig:CFQH}
		\end{center} 
	\end{figure}
	
	In the left panels of figure \ref{fig:CFQH}, $\mathcal{C}$ is plotted versus temperature for both systems. When $\mathcal{C} > 0$, the peak occurs when $\Delta\mathcal{S}$ is activated most rapidly at low temperature. This can be interpreted as a Schottky peak for our system at low temperature \cite{babanli2022,alshorman2018}. In the other hand, when $\mathcal{C} < 0$, the peak indicates that $\Delta	\mathcal{S}$ is decreasing with temperature. The negative sign of $\mathcal{C}$ characterizes a depletion of the transport related entropy contribution with increasing temperature. 
	
	Stronger backscattering or eventually low effective transmission produces a more pronounced peaks in $\mathcal{C}$. This is due to the fact that stronger backscattering produces a stronger energy dependance of $\mathcal{T}(\omega)$ and so a stronger entropy variation $\Delta\mathcal{S}$. 
	
	In the right panels of figure \ref{fig:CFQH}, thermoelectric transport heat capacity density is plotted versus Fermi energy for both systems. Near the central symmetry region $\varepsilon_F \approx 0$, $\mathcal{T}(\omega)$ changes rapidly with energy which gives a strong thermoelectric response, and thus a maximum $\Delta	\mathcal{S}$ (a minimum $\mathcal{C}$). The positive and negative peaks can be interpreted as the thermodynamic response to shifting $\varepsilon_F$.
	
	Using the approximation given in Eq. \eqref{eq:Sapp}, the thermoelectric transport heat capacity density at low temperature limit, take a Mott's-like formula form, and is given by:
	\begin{equation}
		\mathcal{C}\xrightarrow[]{T\to0} \frac{\pi eV k_B^2 T_0}{3 v_F \hbar}\frac{\partial\ln(\mathcal{T}(\varepsilon))}{\partial \varepsilon}\bigg|_{\varepsilon=\varepsilon_F},
	\end{equation}
	It is clear that, our expression in low temperature regime shows a linear dependence on $T_0$ (similar to the entropy \eqref{eq:EntropyLowT}), which is similar to classical $1D$ electron Gas \cite{kane1997,johnston2020}, $C=\alpha T_0$
	where the proportional factor in the standard condensed matter physics is given by $\alpha=\frac{\pi  k_B^2 }{3\hbar v_F}$ \cite{johnston2020}, while in our model this factor is $\alpha_{\text{Present work}}=\frac{\pi eVL k_B^2 T_0}{3 v_F \hbar}\frac{\partial\ln(\mathcal{T}(\varepsilon))}{\partial \varepsilon}\big|_{\varepsilon=\varepsilon_F}$. Now we can get a constraint on the backscattering amplitude $\tilde{A}$ to get the same expression as the standard result. Using the transmission coefficient $\mathcal{T}(\omega)$ given by Eq. \eqref{eq:Tw}, we find:
	\begin{equation}
		\alpha_{\text{Present work}}=\alpha\frac{eV}{\varepsilon_F} \bigg(\frac{2\tilde{A}^2}{4\varepsilon_F^2+\tilde{A}^2}\bigg),
	\end{equation}
	From this one we conclude that our model is equal to the standard results when we impose the following constrain on the backscattering amplitude $\tilde{A}$, where this value enable us to converge to $\alpha$, which is valid when $\tilde{A}\simeq2\varepsilon_F$, and for weak applied voltage $eV\to\varepsilon_F$, and these lead to $\alpha_{\text{Present work}}=\alpha$.

	%----------------------------------------------------------------------------
	%----------------------------------------------------------------------------

	\section{Conclusion} \label{conc}
	
	In this work, we have investigated the nonlinear thermopower and linear Seebeck coefficient of a TLL with an impurity and interaction parameter $K=1/2$. Exact analytical expressions were derived within the framework of bosonization and refermionization techniques. In the low-temperature regime and for specific Fermi energy scales, the Seebeck coefficient was shown to be directly proportional to the entropy per charge carrier. This equivalence allowed us to apply the Kelvin formula to derive the entropy density variation $\Delta \mathcal{S}$ associated with thermoelectric transport, as well as the charge-carrier contribution to the heat capacity density $\mathcal{C}$. These theoretical results were analyzed for two distinct physical realizations: a quantum point contact (QPC) between a fractional quantum Hall edge state ($\nu=1/3$) and a normal metal ($\nu=1$), and a one-channel quantum conductor coupled to an Ohmic environment (OCC).
	
	The behavior of the nonlinear thermopower $Q$ exhibits a distinct peak structure at low temperatures, signaling a thermoelectric enhancement driven by energy filtering of transmitted charge carriers and the strong energy dependence of the transmission coefficient. The voltage dependence reflects a clear breakdown of electron-hole symmetry. Thermopower production becomes significantly more pronounced in the strong-backscattering regime, where the effective transmission is suppressed. In the linear-response regime, the Seebeck coefficient $S_{\text{Seebeck}}$ exhibits similar characteristics; however, its maximum magnitude is smaller than in the nonlinear case because $Q$ integrates transmission over an extended voltage-driven energy window rather than relying solely on the local energy derivative at the Fermi energy. In both regimes, $Q$ and $S_{\text{Seebeck}}$ obey Mott-like relations at low temperatures.
	
	By applying the Kelvin formula, we derived the variation of the entropy density $\Delta \mathcal{S}$ carried by the transmitted charge carriers. Strong backscattering enhances the energy selectivity of the transmission, producing a maximum in the entropy density at low temperatures. This rapid variation in $\Delta \mathcal{S}$ gives rise to a pronounced peak in the heat capacity density $\mathcal{C}$. Both the entropy variation density and the heat capacity density associated with thermoelectric transport follow Mott-like expressions in the low-temperature limit.
	
	Overall, our findings highlight the dual role of backscattering in 1D quantum systems: while it suppresses electrical conductance, it dramatically enhances non-equilibrium thermopower and amplifies local thermodynamic signatures. Comparing the two physical setups reveals that while the low-temperature regime is characterized by robust Mott-like power-law scaling, increasing the temperature drives a transition toward non-monotonic behavior in both the QPC and OCC systems. The present work established here provides a unified picture connecting non-equilibrium quantum transport to mesoscopic thermodynamics, offering valuable insights for probing and designing energy-filtering quantum thermoelectric devices.

	\acknowledgments

\bibliographystyle{unsrt}

\bibliography{ref}

\end{document}